
\input harvmac
\noblackbox

\def\ie{{\it i.e.} }
\def\bfone{\relax{\rm 1\kern-.35em 1}}
\def\inbar{\vrule height1.5ex width.4pt depth0pt}

\def\IC{\relax\,\hbox{$\inbar\kern-.3em{\rm C}$}}
\def\ID{\relax{\rm I\kern-.18em D}}
\def\IF{\relax{\rm I\kern-.18em F}}
\def\IH{\relax{\rm I\kern-.18em H}}
\def\II{\relax{\rm I\kern-.17em I}}
\def\IN{\relax{\rm I\kern-.18em N}}
\def\IP{\relax{\rm I\kern-.18em P}}
\def\IQ{\relax\,\hbox{$\inbar\kern-.3em{\rm Q}$}}
\def\us#1{\underline{#1}}
\def\IR{\relax{\rm I\kern-.18em R}}
\font\cmss=cmss10 \font\cmsss=cmss10 at 7pt
\def\ZZ{\relax\ifmmode\mathchoice
{\hbox{\cmss Z\kern-.4em Z}}{\hbox{\cmss Z\kern-.4em Z}}
{\lower.9pt\hbox{\cmsss Z\kern-.4em Z}}
{\lower1.2pt\hbox{\cmsss Z\kern-.4em Z}}\else{\cmss Z\kern-.4em
Z}\fi}
\def\a{\alpha}  
 \def\c{\gamma}
\def\G{\Gamma} 
 
 \def\cB{{\cal B}}

 \def\cM{{\cal M}}
\def\cN{{\cal N}}

\def\om{\omega}
\def\vt{\vartheta}
\def\nup#1({Nucl.\ Phys.\ $\us {B#1}$\ (}
\def\plt#1({Phys.\ Lett.\ $\us  {B#1}$\ (}
\def\cmp#1({Comm.\ Math.\ Phys.\ $\us  {#1}$\ (}
\def\prp#1({Phys.\ Rep.\ $\us  {#1}$\ (}
\def\prl#1({Phys.\ Rev.\ Lett.\ $\us  {#1}$\ (}
\def\prv#1({Phys.\ Rev.\ $\us  {#1}$\ (}
\def\mpl#1({Mod.\ Phys.\ Let.\ $\us  {A#1}$\ (}
\def\ijmp#1({Int.\ J.\ Mod.\ Phys.\ $\us{A#1}$\ (}
\def\jag#1({Jour.\ Alg.\ Geom.\ $\us {#1}$\ (}
\def\tit#1|{{\it #1},\ }

\def\Coe#1.#2.{{#1\over #2}}
\def\coeff#1#2{\relax{\textstyle {#1 \over #2}}\displaystyle}
\def\coe#1.#2.{\relax{\textstyle {#1 \over #2}}\displaystyle}
\def\half{{1 \over 2}}

\def\tE{\widetilde E}
\def\barE{\overline{E}}
\def\hE{\widehat E}
\def\ttau{\widetilde \tau}
\def\tq{\widetilde q}

%
\lref\KMV{A.\ Klemm, P.\ Mayr and C.\ Vafa,
{\it BPS states of exceptional non-critical strings,} to appear
in the proceedings of the conference {\it Advanced Quantum
Field Theory,} (in memory of Claude Itzykson)
CERN-TH-96-184, hep-th/9607139.}
\lref\LMW{W.~Lerche, P.~Mayr and N.P.~Warner,
{\it Non-Critical Strings, Del Pezzo Singularities
and Seiberg-Witten Curves,} \nup{499} (1997) 125;
hep-th/9612085.}
\lref\MNWa{J.~Minahan, D.~Nemeschansky and
N.P.~Warner, {\it Investigating the BPS Spectrum of Non-Critical
$E_n$ Strings,} \nup{508} (1997) 64; hep-th/9705237.}
\lref\MNWb{J.~Minahan, D.~Nemeschansky and
N.P.~Warner, {\it Partition Functions for the BPS States of
the $E_8$ Non-Critical String,} USC-97/009, NSF-ITP-97-091,
hep-th/9707149.}
\lref\MNWc{J.~Minahan, D.~Nemeschansky and
N.P.~Warner, {\it Instanton Expansions for Mass Deformed $N=4$
Super Yang-Mills Theories,} USC-97/016
hep-th/9710146.}
\lref\GMS{O.\ Ganor, D.\ Morrison and N.\ Seiberg,
\nup{487} (1997) 93,  hep-th/9610251.}
\lref\Claude{C.~Itzykson, \ijmp{B8} (1994) 1994.}
\lref\AStrom{ A. \ Strominger {\it  Open P-Branes},
\plt{383} (1996) 44, hep-th/9512059.}
\lref\tHo{ G. \ 't Hooft , {\it  On the Phase Transition Towards
Permanent Quark Confinement}, \nup{138} (1978) 1;
G. \ 't Hooft , {\it  A Property of Electric and Magnetic Flux in
Non-Abelian Gauge Theories}, \nup {153} (1979) 141.}
\lref\DMCV{D. \ Morrison and C.\ Vafa,
{\it Compactifications of F-Theory on Calabi-Yau Threefolds  $I$,}
\nup {473} (1996) 74, hep-th/9602114;
 {\it Compactifications of F-Theory on Calabi-Yau Threefolds  $II$,}
\nup {476} (1996) 437, hep-th/9603161. }
\lref\CVFT{ C.\ Vafa, {\it  Evidence for F-Theory,}
\nup{469} (1996) 403, hep-th/9602022.}
\lref\BSV{M. \ Bershadsky, V.  \ Sadov and C.\ Vafa, {\it  D-Branes
and Topological Field Theories} \nup {463} (1996) 420,  
hep-th/9511222.}
\lref\OGanE{O.J.~Ganor, {\it A Test of The Chiral E8 Current
Algebra on a 6D Non-Critical String,} \nup{479} (1996) 197;
hep-th/9607020.}
\lref\CVEW{C.\ Vafa and E. \ Witten,
{\it  A Strong Coupling Test of S-Duality}
\nup {431} (1994) 3, hep-th/9408074.}
\lref\HOCV{H. \ Ooguri and C.\ Vafa, {\it  Two-Dimensional Black
Hole and Singularities of CY Manifolds} \nup {463} (1996) 55,
hep-th/9511164.}
\lref\KatzMV{ S. \ Katz, P.  \ Mayr and  C.\ Vafa,
{\it  Mirror symmetry and Exact Solution of $4D$ $N=2$ Gauge
Theories $I$}, hep-th/9706110 .}
\lref\KKV{S. \ Katz, A. Klemm and C.\ Vafa, {\it Geometric
Engineering of Quantum Field Theories}, \nup{497} (1997) 173,
hep-th/9609239.}
\lref\CVinst{ C. \ Vafa, {\it  Instantons on D-branes}, \nup{463}
(1996) 435,  hep-th/9512078.}
\lref\MDbrane{ M. \ Douglas, {\it  Branes within Branes},
hep-th/9512077.}
\lref\EWMandF{ E. \ Witten,
{\it  Phase Transitions in M-Theory and F-Theory},
 \nup {471} (1996) 195,  hep-th/9603150.}
\lref\GMEW{G. \ Moore and E. \ Witten,
{\it Integration over the u-plane in Donaldson theory},   
hep-th/9709193.}
\lref\MDVV{R. \ Dijkgraaf, G. \ Moore, E. \ Verlinde and  H. \  
Verlinde
{\it  Elliptic Genera of Symmetric Products and Second Quantized  
Strings},
Commun.Math.Phys. 185 (1997) 197,  hep-th/9608096 .}
\lref\MBCV{ M. \ Bershadsky and C. \ Vafa, {\it  Global Anomalies
and Geometric Engineering of  Critical Theories in Six Dimensions},
hep-th/9703167 .}
\lref\VWtor{ C. \ Vafa and E. \ Witten, {\it  On Orbifolds with
Discrete Torsion}, J. Geom. Phys.15 (1995) 189,  hep-th/9409188.}
\lref\SW{N.\ Seiberg and E.\ Witten, \nup426(1994) 19,
hep-th/9407087; \nup431(1994) 484, hep-th/9408099.}
\lref\SWP{N.\ Seiberg and E.\ Witten, \nup{471} (1996) 121,
hep-th/9603003 }
\lref\CFIV{S. \ Cecotti, P. \ Fendley, K.\ Intriligator and  

C. \ Vafa, {\it A New Supersymmetric Index,} \nup{386} 

(1992) 405, hep-th/9204102.}
\lref\ASCV{A.~Strominger and C.~Vafa, \plt{379} (1996) 99,
 hep-th/9601029.}
\lref\MSW{J.~Maldacena, A. ~Strominger and E.~Witten,
{\it Black Hole Entropy in M-Theory}, hep-th/9711053.}
\lref\CVBH{ C. Vafa, {\it Black Holes and Calabi-Yau Three-folds,}
 HUTP-97-A066, hep-th/9711067.}
\lref\EWcom{E.~Witten, {\it Some Comments on String Dynamics,}
IASSNS-HEP-95-63, hep-th/9507121; in {\it Strings `95: Future
Perspectives in String  Theory,} Editors: I.~Bars, P.~Bouwknegt,
J.~Minahan, D.~Nemeschansky, K.~Pilch, H.~Saleur and N.P.~Warner,
Proceedings of USC conference, Los Angeles, March 13-18, 1995,
World Scientific, Singapore, (1996). }
\lref\OGAH{O.~Ganor and A.~Hanany,  \nup{474} (1996) 122,
hep-th/9602120.}
\lref\BCOV{M.~Bershadsky, S.~Cecotti, H.~Ooguri and C.~Vafa,
\nup{405} (1993) 279, hep-th/9302103; \cmp{165} (1994) 311,
hep-th/9309140.}
\lref\SLIMF{S.~Lang, {\it Introduction to Modular Forms,}
Springer-Verlag, (1976).}
\lref\STY{S.-T.~Yau, {\it Essays on Mirror Manifolds,}
International Press, (1992).}
\lref\DKV{M.R.~Douglas, S.~Katz and C.~Vafa, \nup{497} (1997) 155,
hep-th/9609071 }
\lref\DMNS{D.R.~Morrison and N.~Seiberg,\nup{483} (1997) 229,
 hep-th/9609070 }
\lref\VWmath{K. Yoshioka, {\it The Betti Numbers of The Moduli Space
of Stable Sheaves of Rank 2 on ${\bf P}^2$,} preprint, Kyoto
University; \hfil \break
K. Yoshioka, {\it The Betti Numbers of The Moduli Space
of Stable Sheaves of Rank 2 on a Ruled Surface,} preprint,
Kyoto University; \hfil \break
A. A. Klyachko,  {\it Moduli of Vector Bundles and Numbers
of Classes,} Funct. Anal. and Appl. {\bf 25} (1991) 67.}
\lref\LeVa{N.~C.~Leung and  C. \ Vafa, {\it  Branes and Toric
Geometry}, hep-th/9711013.}
\lref\EWsix{E.~Witten, {\it Physical Interpretation Of Certain Strong
Coupling Singularities}, \mpl{11} (1996) 2649, hep-th/9609159.}
\lref\EVerl{E.~Verlinde, {\it Global Aspects of Electric-Magnetic
Duality,} \nup{455} (1995) 211; hep-th/9506011.}
\lref\OGan{O.J.~Ganor, {\it Compactification of Tensionless String  
Theories,}
hep-th/9607092.}
\lref\Nekr{ A. Lossev, N. Nekrassov and S. Shatashvili,
{\it Testing Seiberg-Witten Solution}, hep-th/9801061.}
\lref\FMW{R~Friedman, J.W.~Morgan and E.~Witten, {\it Vector
Bundles over Elliptic Fibrations}, alg-geom/9709029.}
\lref\BJPS{M.~Bershadsky, A.~Johansen, T.~Pantev and V.~Sadov,
{\it  On Four-Dimensional Compactifications of F-Theory,}
\nup{505} (1997) 165; hep-th/9701165.}
\lref\Veretal{R. Dijkgraaf, E.
Verlinde and H. Verlinde, \nup{486} (1997) 89.}
\lref\KY{K. Yoshioka, to appear.}
\lref\GZ{L. Gottsche and D. Zagier, {\it 
Jacobi forms and the structure of Donaldson invariants for 4-manifolds with 
$b_+=1$}, alg-geom/9612020.}
%
%
\Title{\vbox{
\hbox{USC-98/002}
\hbox{HUTP/A012}
\hbox{\tt hep-th/9802168}
}}{\vbox{\centerline{\hbox{$E$-Strings and $N=4$ Topological
Yang-Mills Theories}}
\vskip 8 pt
\centerline{ \hbox{}}}}
\centerline{J.A.~Minahan$^*$, D.~Nemeschansky$^*$, C.~ Vafa$^\dagger$
and N.P. Warner$^*$}
\bigskip
\bigskip
\bigskip
\centerline{$^*${\it Physics Department, U.S.C., University Park,
Los Angeles, CA 90089-0484, USA}}
\medskip
\centerline{$^\dagger${\it Lyman Laboratory of Physics,
Harvard University, Cambridge, MA 02138, USA}}
\bigskip
We study certain properties of six-dimensional
tensionless E-strings (arising from zero size $E_8$ instantons).
In particular we show that $n$ E-strings
form a bound string which carries an $E_8$ level $n$ current algebra
as well as a left-over conformal system with $c=12n-4-{248n\over  
n+30}$,
whose characters can be computed.  Moreover we show that the  
characters
of the $n$-string bound state are
 captured by $N=4$ $U(n)$ topological Yang-Mills theory
on $\half K3$.  This relation
not only illuminates certain aspects of E-strings but can also be  
used to shed
light on the properties of $N=4$ topological Yang-Mills theories
on manifolds with $b_2^+=1$.
In particular the E-string partition functions, which can be computed
using local mirror symmetry on a Calabi-Yau three-fold,
give the Euler
characteristics of the Yang-Mills instanton moduli space on $\half  
K3$.
Moreover,
the partition functions are determined by a gap condition
combined with a simple recurrence relation which has its origins in a
holomorphic anomaly that has been conjectured
to exist for $N=4$ topological
Yang-Mills on manifolds with $b_2^+=1$ and is also related
to the holomorphic anomaly for higher genus topological strings
on Calabi-Yau threefolds.

\vskip .3in


\Date{\sl {February, 1998}}

%
\parskip=4pt plus 15pt minus 1pt
\baselineskip=15pt plus 2pt minus 1pt
%
\newsec{Introduction}

One of the interesting new results in string theory has
been the discovery of various kinds of strings
in addition to the fundamental string.  Among these, there
are strings where the tension is large, and these have applications
in the counting of microstates of blackholes \ASCV.  At the other
extreme there are constructions in which the string can become
tensionless, and one typically ends up with a critical conformal
theory.  There are several  natural questions one could ask about
the physical properties of such strings:    How are they
different from fundamental strings?  What are their excitations?  Do
tensionless strings bind together?

One of the commonalities between fundamental and non-critical
strings is that, for a sufficient number of compactified dimensions,
all of these strings can be related to M5 branes wrapped around a
four cycle ${\cal N}$ (for applications to black hole counting
in this context see \refs{\MSW,\CVBH}).  For example, the type II
string can be viewed as an M5 brane wrapped around $T^4$, and
the heterotic string can be viewed as an M5 brane wrapped around  
$K3$.
There are two well-known examples of non-critical strings in six
dimensions, one with $(0,2)$ (space-time) supersymmetry \EWcom\
and the other with $(0,1)$ (space-time) supersymmetry.  The second
is dual to the heterotic string
with an $E_8$ instanton of zero size \refs{\OGAH,\SWP}, which in
F-theory corresponds to a  particular ${\bf P}^1$ shrinking  to zero
size \refs{\DMCV,\EWMandF}.  The $(0,2)$ non-critical string can be
viewed as an M5 brane wrapped around ${\bf P}^1\times T^2$, while the
$(0,1)$ non-critical string can be viewed as an  M5 brane wrapped
around $\half K3$.  The purpose of this paper is to continue the
study of the BPS states of  the $(0,1)$ non-critical string,  
initiated
in \KMV\ and further studied in detail in \refs{\MNWa,\MNWb}.
We will call such strings, E-strings, for their association
with the shrinking $E_8$ instanton.  These strings have
less world-sheet supersymmetry, $(0,4)$, in comparison
to the other string, which has $(4,4)$ world-sheet supersymmetry.
The generalized elliptic genus, which counts BPS states
weighted by $ (-1)^F$, vanishes for the $(0,2)$ string, but
it is much more informative for the $(0,1)$ string, and will be
discussed extensively in this paper\foot{Note
that ${\bf P}^1\times T^2$ can be viewed as $\half T^4$ in the sense
that ${\bf P}^1=T^2/Z_2$.  It turns out that the low energy
degrees of freedom on the $(4,4)$ string is exactly half
of the degrees of freedom of the type II string (in the lightcone)
which in turn corresponds to an M5 brane wrapped around $T^4$.
Similarly for the $(0,4)$ case, which corresponds to an M5 brane
wrapped around $\half K3$,
the degrees of freedom of the single E-string is
 half  that of the $E_8\times E_8$ heterotic string,
which in turn corresponds to the M5 brane wrapped around $K3$.
So perhaps another set of suitable names for these strings are
 $\half$ type II strings and $\half$ heterotic strings!}.

The study of the BPS states suggest many new things
about the E-strings.  It shows that $n$ of them form
bound states at threshold and that they carry an $E_8$ level
$n$ Kac-Moody algebra.  Moreover, there are additional left-over
degrees of freedom on the left-movers for which we compute the
corresponding characters.

One of the key observations in this paper is to use the connection
of such BPS counts with the computation of $N=4$ topological
Yang-Mills \CVEW.   
This relation shows,
in particular that the number of BPS states consisting of $n$ wrapped
E-strings with total momentum $k$ around the circle, which is
given by counting holomorphic curves on $\half K3$, is related (by an
application of mirror symmetry) to the partition function of
$N=4$ topological $U(n)$ Yang-Mills on $\half K3$ with
instanton number $k$.  Using this observation we relate
the holomorphic anomaly discovered for $N=4$ $SU(2)$ Yang-Mills
on manifolds with $b_2^+=1$ \CVEW, to the holomorphic
anomaly in the BPS state partition function derived
in \refs{\MNWb}.   In \MNWb\ the holomorphic anomaly was derived for
arbitrary $n$, and it now gives us the holomorphic anomaly of $SU(n)$  
$N=4$
gauge theories.  We will also  give an interpretation of the  
holomorphic
anomaly from the viewpoint of counting holomorphic curves, which
thus connects to the holomorphic anomaly of Kodaira-Spencer theory
\BCOV.  Furthermore we interpret
the existence of this holomorphic anomaly as coming
from the fact that $n$ E-strings form bound states at threshold.

The organization of this paper is as follows:
In section 2 we review various interconnected ways of looking at
the E-strings.  We also relate the BPS count
with the $N=4$ topological partition function.
In section 3 we discuss general aspects of $N=4$ topological
Yang-Mills theory on four manifolds, and we develop the $N=4$ $U(n)$
Yang-Mills theory on $K3$ in detail (generalizing
the results in \CVEW\ from $SU(2)$ to $U(n)$).  We then discuss
aspects of $N=4$ Yang-Mills
on $\half K3$ which follow from general principles and a few  
conjectures.
In section 4 we discuss the general results concerning the BPS state
partition function for E-strings.  In section
5 we review, and re-derive in a new way, the mirror of $\half K3$  
which
is needed for the BPS count.  In section 6 we write very explicitly
the partition function for
low winding numbers (up to $4$).  The results are compared with
the predictions based on $N=4$ Yang-Mills.  In section 7 we try
to interpret some aspects of the BPS partition function as well as
the meaning of the anomaly in terms of the E-string.  We conclude,
in section 8, with some suggestions for further study.

\newsec{F-theory, M-theory, Yang-Mills theory and E-strings}

In this section we will review the various descriptions
of solitonic strings in F-theory and M-theory.  We start with
the $N=1$ supersymmetric compactification of F-theory to six
dimensions, and then we discuss how the same string
looks from the M-theory perspective.  That is,  upon a further
compactification to five dimensions on an $S^1$, such a string
theory has two dual descriptions in M-theory that depend upon whether
the string wraps $S^1$ or not.  Both descriptions are relevant for us
and will be reviewed here. However it is the M5-brane description of
the string that enables us to connect the partition function of the
string with that of topologically twisted $N=4$  supersymmetric
Yang-Mills theory.

\subsec{General overview}

Consider compactification of F-theory down to six dimensions
on an elliptic Calabi-Yau threefold, $K$, with a section, \ie with
a two complex dimensional base sub-manifold $B$ \refs{\CVFT,\DMCV}.
Let $\Sigma \subset B$ denote a Riemann surface in the base.
Let $\hat \Sigma$ denote the
complex two dimensional elliptic manifold consisting of $\Sigma$ and
the elliptic fiber of $K$ over each point.  Consider the 3-brane
of type IIB wrapped around $\Sigma$, which gives rise to a string
in six dimensions.  One is interested in the low energy degrees of
freedom on this string.  One aspect of these degrees of freedom is
reflected in the BPS states that result upon further wrapping of
this string $n$ times around a circle, carrying some momenta $p$.
This gives rise to BPS particles in five dimensions.
{}From the M-theory perspective these are described as follows:
The compactification of F-theory on
$K\times S^1$, where $S^1$ has radius $R$, is equivalent to an
$M$-theory compactification on $K$ in which the K\"ahler class of the
elliptic fiber is $1/R$.  In $M$-Theory, the BPS states
described above correspond
to M2 branes wrapped around $K$ whose $H_2$ class is given by
$n[\Sigma]+p[T^2]$, where $[T^2]$ denotes the class of the elliptic
fiber.  Thus counting the BPS states of the six-dimensional string
wrapped around the circle amounts to counting holomorphic curves on
$K$, which in turn can be done using mirror symmetry techniques.
This was the method exploited in \KMV\ and will
be discussed further in sections 4 and 5.

The counting of these five-dimensional BPS states can be related to
another computation. Consider a string that remains a string in
five dimensions, that is a string that does not wrap the circle.
In terms of M-theory this is described by wrapping an M5 brane
around $\hat \Sigma$.  Now compactify one more dimension
on a circle of radius $R'$ and wrap this string around the
new circle so that the M5 brane is wrapped around
$\hat\Sigma\times S^1$. The BPS states of this wrapped state can
be computed by its partition function, \ie by taking the
world-volume of the M5 brane to be
$\hat \Sigma \times T^2$.  Moreover if we are interested
in strings wound $n$ times around the second $S^1$  we should
consider $n$ M5 branes whose world-volume is given by $\hat \Sigma
\times T^2$.  However as is well known $n$ M5 branes wrapped around
$T^2$ give rise to $N=4$ $U(n)$ Yang-Mills in four dimensions
\refs{\EWcom,\HOCV,\AStrom},  (where the momentum number $p$
gets mapped to the instanton number).  We are thus left with the
partition function of an $N=4$ Yang-Mills theory on $\hat \Sigma$.
More precisely we end up with a topologically twisted version of
$N=4$, as is expected on general grounds \BSV.  In the next section
we show that the relevant twist is the one already considered
in \CVEW.

Putting this all together, we are relating the number of holomorphic
curves\foot{More precisely, an appropriate Euler characteristic for
their moduli spaces.} wrapped around $n[\Sigma]+p[T^2]$ to the
partition function of $N=4$ $U(n)$ Yang-Mills theory on $\hat  
\Sigma$.
This final result has a relatively simple alternative explanation
in terms of the type IIA description of the same
states coming from a further compactification on a circle:
Consider $n$ D4 branes wrapped around $\hat \Sigma$ and consider
the number of cohomology elements in the instanton moduli space with
instanton number $p$.  This is the same as the number of bound
states of $n$ D4 branes with $p$ D0 branes \refs{\MDbrane,\CVinst}.
Now, recall that $\hat \Sigma$ is elliptic and we are considering
the limit where the size of the elliptic fiber is small.
In this limit we can go to the T-dual description where the
D4 branes become $n$ D2 branes wrapped around $[\Sigma]$
and the D0 branes become $p$ D2 branes wrapped around $[T^2]$.
For BPS states we are interested in their bound states, which
are represented by a holomorphic curve in $\hat \Sigma$ in the
class $n[\Sigma]+p[T^2]$.  Which thus brings us to the same
conclusion via a more direct argument.
In fact this T-duality, relating rank $n$-bundles on elliptic
manifolds to holomorphic curves is also well known mathematically
and is known as the spectral curve and has been studied
in the context of F-theory in  \refs{\FMW,\BJPS}.

We can take this result one step further by a second application
of mirror symmetry.  Namely, we can use (local version of) mirror  
symmetry
\refs{\KMV,\LMW,\KKV}
to count the number of holomorphic curves in $\hat \Sigma$.  We thus
conclude that {\it the partition function of topologically-twisted}
$N=4, U(n)$ {\it Yang-Mills on an elliptic four-dimensional manifold}
$\hat \Sigma$, {\it can be computed by a double application of mirror
symmetry}.   {\it Moreover, this partition function counts the BPS
states of} $n$ {\it times wrapped strings obtained in M-theory by
wrapping M5 branes around} $\hat \Sigma$ {\it
or in F-theory of D3 branes wrapped around} $\Sigma$.

\subsec{Tensionless E-strings from F-theory}

In the foregoing description we have described how one gets a string
in six dimensions for every holomorphic curve $\Sigma \subset B$
in an F-theory compactification on a 3-fold.  However to get an
interesting
critical theory the resulting string will have to be tensionless.
Since the tension of the string is proportional to the volume of
$\Sigma$, a tensionless string will arise only if that $\Sigma$
is shrinkable.  This can only happen if $\Sigma$ is a ${\bf P}^1$.
Moreover the normal bundle of ${\bf P}^1$ in the base $B$
should be negative \DMCV.  Denote its normal bundle in $B$ by
$O(-k)$.
The fact that $c_1$ of the 3-fold is zero implies that
the tangent direction on the elliptic fiber over ${\bf P}^1$ will
then correspond to an $O(k-2)$ bundle.  We restrict our attention
to the situation where the elliptic fibration is generically
non-singular. (Singular fibrations are also interesting and
have also been considered in the literature
\refs{\DMCV,\EWsix,\MBCV}).
Non-singularity requires that the cotangent
bundle of the elliptic fiber have a section over ${\bf P}^1$ which
can happen only if $k=1,2$.  Before discussing these two cases,
we note that the choice, $k=0$, which is not
shrinkable at finite distance in moduli, corresponds to ${\hat
\Sigma}=K3$,
and the corresponding string is the heterotic string.

For $k=2$, the elliptic fibration is trivial over ${\bf P}^1$,
\ie, ${\hat \Sigma}={\bf P}^1\times T^2$, and we get the tensionless
string corresponding to $(4,4)$ susy on the worldsheet (and is the
same as the tensionless string that appears for type IIB theory near
an $A_1$ singularity \EWcom ).  The choice $k=1$ corresponds to
${\hat \Sigma}=
\half K3$.   The reason for this terminology is that, one gets $K3$
as the elliptic fibration over ${\bf P}^1$ with $k=0$, and then
the modulus of the torus covers the fundamental domain $24$ times
(corresponding to $24$ cosmic strings), the $\half K3$ corresponds
to an elliptic manifold over ${\bf P}^1$ whose modulus covers the
fundamental domain $12$ times. More explicitly, it can be described
as
\eqn\weier{y^2 ~=~ x^3 + f(z)x + g(z) \ ,}
where $f$ and $g$ denote polynomials of degree $4$ and $6$ in $z$
respectively,
where $z$ parameterizes the
${\bf P}^1$.  The elliptic fiber
is thus described  by $(x,y)$ subject to the equation \weier.
The weights are such that $y$ belongs to an $O(3)$ bundle
and $x$ to an $O(2)$ bundle on ${\bf P}^1$, and so
the cotangent bundle of the elliptic fiber is an $O(1)$ bundle
$[dy/x]$
over ${\bf P}^1$.

The non-zero hodge numbers for this manifold are
$$h^{0,0}=h^{2,2}=1 \ , \qquad h^{1,1}=10$$
In particular this implies that $b_2^+ = 1$ (for a K\"ahler
manifold of complex dimension $2$ we have $b_2^+=1+2 h^{2,0}$).
This manifold can also be obtained by blowing up ${\bf P}^2$ at
$9$ points (the position of the ninth point is determined by
the other eight points).  For this reason the manifold $\half K3$
is a del Pezzo surface
sometimes denoted by $\cB_9$, and we will use this notation
interchangeably.

We will call the resulting string the E-string.  The reason for this
terminology is that the E-string arises in a dual description  of an
$E_8$ instanton in heterotic string theory
shrinking to zero size.  This connection is explained in
\refs{\EWMandF,\DMCV}.

Since the $E$-string constructed here lives in six-dimensions, and
has  $(0,4)$ world-sheet supersymmetry, it is natural
to define the following generalized form of the elliptic genus
for such a string:
\eqn\zpartf{Z~=~{\tau_2^2 \over V_4} {\rm Tr}~\Big[~(-1)^{F}~F_R^2~
q^{L_0}~ {\overline q}^{\overline{L_0}}~\Big] \ .}
Here we are imagining computing the partition function on
a torus with $q={\rm exp}(2\pi i\tau)$.  The factor of
$F_R^2$ has been inserted to soak up the fermion zero-modes,
and we have divided by the volume, $V_4$, of the four non-compact
bosons representing transverse position of the string.
We have also multiplied by a factor of $\tau_2^2$, where
$\tau_2 = Im(\tau)$, so as to cancel the factor of
$1/\tau_2^2$ coming from the integration over the momenta
of the string in the transverse directions.
This partition function, $Z$, has (holomorphic,
anti-holomorphic) modular weight $(-2,0)$:  Without the factor
of $\tau_2^2$ and without the insertion of $F_R^2$, $Z$ would be
a modular invariant.   Each factor of
$F_R$ increases the anti-holomorphic weight by one unit, and
each factor of $\tau_2$ shifts the weight $(-1,-1)$.  We will
discuss this partition function more extensively below.

\subsec{Yang-Mills partition functions from M5-branes}

Consider $n$ parallel $M5$ branes whose world-volume
is given by a six-dimensional sub-manifold of spacetime
$$\cM^6 ~=~ T^2\times \cN^4$$
As is well known, compactification of $n$ parallel $M5$ branes
on $T^2$ yields a $U(n)$ $N=4$ supersymmetric Yang-Mills on
the left-over world-volume, where the complex structure $\tau$
of $T^2$ gets identified with the complexified $U(n)$ Yang-Mills
coupling $\tau ={4\pi i\over g^2}+{\theta\over2\pi}$.
In this description, the Montonen-Olive
duality is a consequence of the classical $SL(2,{\bf Z})$ symmetry of
$T^2$ \refs{\EVerl,\EWcom}.

This relation with $N=4$ Yang-Mills implies that
the partition function of $n$ five-branes on $T^2\times\cN^4$ is the
same as that of $U(n)$ N=4 Yang-Mills on $\cN^4$.  We can also view
this slightly differently:  We can think of $n$ parallel five-branes
wrapped around $\cN^4$ as giving rise to a string, and then the
partition function of this theory is the same as the partition
function of the effective $(1+1)$-dimensional
theory on the $T^2$ with modular parameter $\tau$.  This viewpoint
has been suggested before in \OGan.

Depending upon the situation in which this arises in the string
theory, one typically ends up with a topologically
twisted version of $N=4$ Yang-Mills theory \BSV .
There are three possible twistings for $N=4$ \CVEW , and we need to
determine
the relevant one here.  For us the $\cN^4$ is embedded in a
Calabi-Yau 3-fold, and the five scalars of the five-brane, correspond
to normal deformations of $\cN^4$.  The normal bundle will be
trivial only in the uncompactified directions (\ie the normal
directions
that are {\it not} in the Calabi-Yau) and so three of the scalars
must continue to transform as scalars after twisting.
The other two will be a section of the canonical bundle on $\cN^4$,
\ie a section of the form $ f(z_1,z_2,{\overline z_1},{\overline
z}_2)
dz_1\wedge dz_2$.  The last scalar of Yang-Mills in four dimensions
comes from giving the anti-symmetric $2$-form of the five-brane
worldvolume an expectation value on the $T^2$.  The result is thus
a periodic scalar. The three uncompactified scalars correspond to
the transversal position of the left-over string of the
M5 branes in uncompactified five-dimensional spacetime.
This uniquely fixes the twisting of the $N=4$ Yang-Mills and is the
one already studied in detail in \CVEW .  This also fixes the
supersymmetry on the effective $(1+1)$-dimensional theory to be
generically $(0,4)$ (for $K3$ the supersymmetry is  $(0,8)$).

Strictly speaking, in such a supersymmetric theory the partition
function of $n$ parallel M5 branes on $T^2\times \cN^4$ is zero.
This is because of the supersymmetric degree of freedom associated
with the translations in the uncompactified space.  Equivalently,
in the Yang-Mills language, this is related to the zero modes coming
from the $U(1)\subset U(n)$. Suppressing translations, by absorbing
the correponding fermion zero modes, will result generically in a
non-vanishing partition function.  In the $(1+1)$-dimensional,
effective theory, this means that we really need to compute the
generalized form of the elliptic genus described earlier:
\eqn\partfs{Z~=~{\tau_2^{3/2} \over V_3} {\rm Tr}~\Big[~
(-1)^{F}~F_R^2~ q^{L_0}~ {\overline q}^{\overline{L_0}}~\Big] \ ,}
where $q={\rm exp}(2\pi i\tau)$.  Once again we divide by the
volume of the space tranverse to the string, and introduce
factors of $\tau_2$ to cancel those coming from
the integration of the corresponding momenta. This transverse
space is three-dimensional since we now have the $E$-string in
five-dimensions, or equivalently, only three of the five scalars
of the M5-brane are non-compact.
This partition function, $Z$, is expected to be a modular form of
(holomorphic,anti-holomorphic) weight $(-3/2,1/2)$ (which
is $(0,2)$ shifted by $(-3/2,-3/2)$ coming from the factors of
$\tau_2$).
If the corresponding $(1+1)$-dimensional theory has a discrete
spectrum then one can argue that the only $\overline \tau$
dependence of $Z$ comes from the bosonic zero modes.   This is
because the $N=4$ supersymmetry for the right-movers implies that
the right-moving oscillator contributions cancel.
In the gauge language these bosonic zero-modes are related
to the bosonic modes in $U(1)\subset U(n)$.
In particular the $SU(n)$ part of the partition function is
expected to be a function of $\tau$ only.

For $K3$ one has eight right-moving bosons and eight right-moving  
fermions.
This means that the generalized elliptic genus \partfs\ has to have
an insertion of $F_R^4$ (as opposed to $F_R^2$) to absorb the
fermion zero-modes.    Moreover, in the compactification of
$M$-theory on $K3$ one is in seven dimensions, with five dimensions
tranverse to the string, and hence one has to divide
by the volume, $V_5$,  of this transverse space. Following the
practice above, we will also multiply by $\tau_2^{5/2}$ to
compensate for the transverse momentum integrations, and so
the partition function $Z$ then has a weight
of $(0,4)+(-5/2,-5/2) = (-5/2,3/2)$.

As discussed above, the function $Z$ has another
interpretation: it is the partition
function of the BPS states obtained by wrapping a string
coming from $n$ parallel M5-branes on $\cN^4$ around another circle.
In this interpretation the coefficients of $q^p$ in $Z$ count the BPS
states which have momentum $p$ around the circle.  From the
Yang-Mills point of view, $p$ is the same as the instanton number
(up to a fixed shift of $-n{\chi (\cN^4)/24}$). Moreover,
for $n$ parallel M5 branes the resulting string can be viewed
as bound states of $n$ singly wound strings around the circle.

\newsec{The partition functions for $N=4$ Yang-Mills}

The partition function, $Z$, of topologically twisted
$N=4$ Yang-Mills was studied for various manifolds in \CVEW.  The
most concrete results were for an $SU(2)$ gauge group on K\"ahler
manifolds with $b_2^+ >1$, where $b_2^+$ denotes the dimension
of the self-dual 2-forms.  Moreover,
general considerations led to certain predictions for all $SU(n)$ on
these manifolds.  The more difficult problem of $SU(2)$ gauge theory
on ${\bf P}^2$ was also considered in \CVEW.  The difficulty lies
in the fact that $b_2^+=1$, and so there are no holomorphic
deformations of the canonical bundle. This means that one cannot
deform the theory  to give masses to the adjoint fields in
a manner that preserves $N=1$ supersymmtery.  We will review $N=4$
Yang-Mills on ${\bf P}^2$ later in this section and also give a
more detailed discussion of Yang-Mills theories and M5-branes on
manifolds lacking  holomorhic deformations of their canonical  
bundles.

Evidence was provided in \CVEW\ that for an $SU(n)$ gauge group on
an arbitrary four manifold, $\cN^4$, with $b_2^+>1$,
$Z$ is a modular form of an $SL(2,Z)$ subgroup
with (holomorphic, anti-holomorphic) weight $(-\chi/2,0)$, where
$\chi$ denotes the Euler characteristic of $\cN^4$. The partition
function
$Z$ has an expansion of the form
\eqn\inexp{Z ~=~q^{-{n\chi\over 24}}~\sum_{k=0}^{\infty}~c_k~q^k}
where $c_k$ denotes the contribution of the instanton number $k$
sector to the partition function.  The overall shift of
$q^{-{n\chi\over 24}}$ is needed to make $Z$ a modular form.
It was also noted in \CVEW\ that when the manifold is ${\bf P}^2$
(and conjectured more generally for K\"ahler manifolds with  
$b_2^+=1$),
$Z$ can only be made modular by adding $\overline \tau$
dependence to $Z$.  Having done this, the expansion above is
valid if one considers $\tau$ and $\overline \tau$ as formal
variables, and $\tau$ is kept fixed while ${\overline \tau}
\rightarrow \infty$.

The partition functions for $SU(n)/Z_n$ gauge groups were also
studied in \CVEW.  For such quotients we can also have non-trivial
`t Hooft fluxes through the two cycles of the manifold \tHo.
Hence, one can write a partition function
$Z_{\alpha}$ for each `t Hooft flux $\alpha \in H^2(M,Z_n)$.
Furthermore, for each $Z_\alpha$ we get an expansion of the form
\inexp, except that the instanton numbers are shifted by
${\alpha^2\over 2n}-{\alpha^2\over 2}$, and are hence generically
fractional. It was also shown  that under $\tau\rightarrow -1/\tau$
the $Z_\alpha$ mix according to
\eqn\shar{Z_\alpha \rightarrow \pm
n^{-b_2/2}(\tau/i)^{-\chi/2}\sum_\beta
{\rm exp}\left({2\pi i \alpha \cdot \beta \over n}\right)Z_\beta \ ,}
where $b_2$ denotes the second betti number of $\cN^4$.
In the next sub-section we discuss some of the results of
\CVEW\ for $K3$.  This will set the stage
for the generalization to ${1\over 2}K3$, which is the main
focus of this paper.

\subsec{Partition functions for $K3$}

 We begin by describing the result for $K3$, with gauge
group $SU(2)$, and we will generalize the result of
\CVEW\ by considering a $U(2)$ gauge theory.
This is straightforward as it merely involves the addition of a
trivially computable free $U(1)$ theory.
In fact, given that we have already suppressed the fermionic degrees
of freedom, this only includes the volume factor from bosons,
together with the $U(1)$ fluxes through 2-cycles. In particular
we can view this theory as
$${SU(2)\times U(1)\over Z_2} \ ,$$
and so we just have to combine the partition functions of $SU(2)/Z_2$
corresponding to various `t Hooft fluxes with the corresponding
$U(1)$ fluxes.

To get started we first consider
the $U(1)$ theory by itself, which corresponds
to a single five-brane.  The self-dual field strength
antisymmetric field
$B_{ij}$ on the five-brane leads to $19$ left-moving and $3$
right-moving bosonic modes, all of which are periodic.  The
three uncompactified scalars give 3 additional left- and
right-movers.  The other two scalars transform according to the
canonical bundle on $K3$, which is in turn trivial,  so this will
give an additional two left- and two right-moving scalars.
All together we have 24 left-moving and 8 right-moving bosonic
degrees of freedom.  Finally, there are also 8 right-moving fermionic
modes.  This
is indeed the oscillator content of the heterotic string.

Since the effective $(1+1)$-dimensional theory has an $(0,8)$
supersymmetry, we need to consider the generalized elliptic genus
\eqn\elgenK{
Z_1~=~{\tau_2^{5/2}\over  V_5}\Tr\left[(-1)^{F_R}{F_R}^4 q^{L_0}\bar
q^{L_0}\right]}
in order to get a non-zero result.
As we remarked earlier, the factor of $\tau_2^{5/2}/V_5$
comes from the five uncompactified bosons transverse to the
string, and the modular weight of $Z$ is $(-5/2,3/2)$.

The result is then made up of two parts:
(i)  A lattice partition function of the $U(1)$ fluxes
through each $2$-cycle, and (ii) the partition function of
the $24$ bosonic zero modes of the single five-brane on
$K3$\Veretal:
$$Z_1~=~ G(q)~\theta_{\Gamma^{19,3}} (q,{\overline q})\ , $$
where
$$G(q)~=~{1\over \eta(q)^{24}}~=~\bigg({1\over q^{1\over 24}\prod_n
(1-q^n)}\bigg)^{24}$$
and
$$\theta_{\Gamma^{19,3}}(q,{\overline q})~=~\sum_{{P_L,P_R}\in
\Gamma^{19,3}}~ q^{\half P_L^2}~{\overline q}^{\half P_R^2} \ .$$
The lattice, $\Gamma^{19,3}$, is the even, self-dual, integral
lattice with signature $(19,3)$ with a choice of polarization
(the standard Narain sum).  The $U(1)$ gauge fluxes of the $N=4$
gauge theory correspond to self-dual $H$-fluxes on the five-brane.
Note that $Z_1$ has the desired modular weight of $(-5/2,3/2)$
with  $(19/2,3/2)$ coming from the $\theta$-function  of the
$\Gamma^{19,3}$  lattice, and  $(-12,0)$ coming from the
$\eta$-functions.

Now we consider the $U(2)$ theory, \ie the theory of two
five-branes wrapped on $K3\times T^2$. The coupling constant
of $U(2)$ relative to that of $U(1)$ is naturally halved  via
$\tau \rightarrow \tau /2$.  This means that if we were considering
the $U(1)$ theory we would  replace $(q,{\overline q})$
with $(q^{\half},{\overline q}^{\half})$ in the
lattice theta function.  However this is not quite correct because
the corresponding fluxes are correlated with the $SU(2)/Z_2$
't Hooft fluxes.  These come in three categories:  trivial, even and
odd.  The trivial flux is correlated with those lattice vectors of
$\Gamma^{19,3}$ that are twice another vector.  The even 't Hooft
fluxes correlate  with $U(1)$ fluxes with $\half(P_L^2-P_R^2)$ even,
but where $(P_L,P_R)$ is not twice another vector.  The odd fluxes
correlate with $U(1)$ fluxes where $\half (P_L^2-P_R^2)$ is odd.
The three pieces of $U(1)$ fluxes correspond to
$$\theta_0=\theta_{\Gamma^{19,3}} (q^2,{\overline q}^2) \ , $$
$$\theta_{even}={\half}(\theta_{\Gamma^{19,3}}(q^{\half},{\overline
q}^{\half})+\theta_{\Gamma^{19,3}}(-q^{\half},-{\overline  
q}^{\half}))
-\theta_{\Gamma^{19,3}} (q^2,{\overline q}^2) \ ,$$
$$\theta_{odd}={\half}(\theta_{\Gamma^{19,3}}(q^{\half},{\overline
q}^{\half})-\theta_{\Gamma^{19,3}}(-q^{\half},
-{\overline q}^{\half})) \ .$$
Note that the sum of these three terms is simply
$\theta_{\Gamma^{19,3}}(q^\half,{\overline q}^\half)$ as it should  
be.
The correlation between $SU(2)/Z_2$ `t Hooft fluxes
and the $U(1)$ fluxes now imply that the partition function of $U(2)$
is given by
$$Z_2=Z_0 \theta_0+Z_{even }\theta_{even}+ Z_{odd}\theta_{odd}$$
where $Z_0,Z_{even}$ and $Z_{odd}$ denote the corresponding
$SU(2)/Z_2$ partition functions.  From the results in \CVEW\
we have
$$\eqalign{Z_0~=~& \coeff{1}{4}G(q^2)+\coeff{1}{2}[G(q^\half)+
G(-q^\half)] \ , \cr
Z_{even}~=~& \coeff{1}{2}[G(q^\half)+G(-q^\half)] \ ,
\qquad Z_{odd}~=~ {1\over 2}[G(q^\half)-G(-q^\half)] \ .}$$
Therefore we obtain:
\eqn\utp{Z_2=\coeff{1}{4}G(q^2)\theta_{\Gamma^{19,3}}
(q^2,{\overline q}^2)+ \coeff{1}{2}[G(q^\half)\theta_{\Gamma^{19,3}}
(q^{\half}, {\overline q}^{\half})+ G(-q^\half)\theta_{\Gamma^{19,3}}
(-q^{\half}, -{\overline q}^{\half})] \ .}
Note that the partition function $Z_2$ is a modular form of
$SL(2,Z)$.
Moreover it has the same weight as $Z_1$, namely it has weight
$(w_L,w_R)=(-{5\over 2},{3\over 2})$.  The fact that it is a modular
form of $SL(2,Z)$ is consistent with Montonen-Olive self-duality
for a $U(2)$ gauge group.

We now try to interpret the $Z_2$ partition function from the
five-brane point of view.  If we have two copies of $K3 \times T^2$
in spacetime then this can be viewed as a single five-brane wrapped
twice
over  $K3\times T^2$.  This can be done in several ways by
taking coverings  over $K3$ or over $T^2$.  As a  hint we can use the
fact that a five-brane wrapped around $K3$ is equivalent to the
heterotic
string.  For two copies of the  heterotic string all one has to do
is  double cover the worldsheet.  This suggests  that
we should consider the two five-brane worldvolume to be
$K3\times T^2$ but with the $T^2$ wrapping twice around the original
spacetime $T^2$. This means that the complex structure of the
torus is now going to be different from that of the $T^2$ in
spacetime, depending on how the world-volume $T^2$ wraps the  
spacetime
$T^2$.  There are three
inequivalent ways in which a $T^2$ can double cover another $T^2$,
resulting in one of the following complex moduli:
\eqn\tauchoices{{\tilde \tau}=2\tau,\ {\tau \over 2},\
{\tau \over 2}+\half \ .}
We would thus expect
\eqn\hkt{Z_2 (\tau) =\coeff{1}{4} Z_1(2 \tau)+ \coeff{1}{2}
Z_1\big ({\tau \over 2}\big)+\coeff{1}{2} Z_1\big(
{\tau \over 2}+\half\big)}
where the constant in front of the $Z_1$ terms is fixed by
modularity,
up to an overall constant of proportionality which can be fixed
by viewing the five-brane partition function as the counting of
BPS states.  This is in clear agreement with \utp .
It is now easy to generalize this to the case of $n$ parallel
fivebranes with worldvolume $K3\times T^2$, which should be
equivalent to $U(n)$ gauge theory on $K3$ (this is similar and  
related
to the observation in \MDVV).  We first have to recall
how to enumerate the inequivalent ways in which a $T^2$ can cover
another $T^2$ $n$ times:  One chooses a basis of $2$-cycles on
$T^2$, and then the inequivalent $n$-fold covers are given by the
$GL(2)$ transformations that can be written in the form
$$
 \left(\matrix{a&b \cr 0&d \cr}\right) \ ,$$
with $ad=n$ and $b< d$ and $a,b,d\geq 0$.  Note that for $n=2$
this gives the three possibilities in \tauchoices.  Thus, taking
into account the modularity of the partition function we find that
\eqn\hecke{Z_n={1 \over n^{2}}~\sum_{a,b,d} ~d ~Z_1
\left({a\tau +b\over d}\right) \ ,}
where the sum is over ${a,b,d}$ satisfying the conditions given
above.  The modular form $Z_n$ in \hecke\ is known as a Hecke
transformation of order $n$ \SLIMF,
which maps a modular form of weight $k$ to another modular
form of weight $k$.  For a general
modular form with holomorphic/anti-holomorphic weights
$(w_L,w_R)$, the powers of $n$ and
$d$ that appear in the Hecke transformation are
$n^{w_L+w_R-1}$ and $d^{-w_L-w_R}$.

This general structure is in agreement with what was conjectured in
\CVEW .  Namely, it was observed that for manifolds with
$b_2^+> 1$ one can deform the Yang-Mills theory by adding
masses to the three chiral
fields, so that locally one  has an $N=1$ theory, without changing
the topological partition function\foot{In
fact $(b_2^+ -1)/2$ is the complex dimension of moduli of holomorphic
deformations of $\cN^4$.  In the five-brane description
the existence of these deformations means
$\cN^4$ can be moved holomorphically in the Calabi-Yau, \ie
the five-branes can be moved off of each other.}.
In terms of $N=1$ chiral fields
the deformed superpotential takes the form
$$W ~=~ Tr[[X,Y],Z] ~-~{m \over 2}~Tr(X^2+Y^2+Z^2)$$
where $X,Y$ and $Z$ denote the chiral superfields in the
adjoint of $SU(n)$.
It was argued  that for $SU(n)$ gauge
theory these deformations lead to a number of inequivalent vacua,
parameterized by how $SU(n)$ is broken in each vacuum.  Namely the
critical points of $W$ are found by solving
$$[X,Y]=mZ, \ \ [Y,Z]=mX,\ \ [Z,X]=mY,$$
which after suitable rescaling, implies that
$X,Y$ and $Z$ form an $n$ dimensional representation of $SU(2)$.
Typically, choosing such vacua results in  gauge groups containing
at least one unbroken $U(1)$.  But these vacua contain extra fermion
zero
modes and so cannot contribute to the partition function.  Hence, we
only need to consider those vacua that have no unbroken $U(1)$
groups.
If $n=ad$,  then this can  be done by choosing the $n$ dimensional
representation to be $d$ copies of the $a$-dimensional irreducible
representation of $SU(2)$.  In this case we find  an $N=1$
theory with $SU(d)$ gauge symmetry.  This will have $d$ vacua, which
we label by $b=0,1,...,d-1$.
Moreover each
vacuum is expected to contribute to the total partition function
and shifting $\tau$ by a constant mixes the different vacua \CVEW.
Thus we find a partition function that
matches the Hecke structure of \hecke .

The Hecke structure of \hecke\ has a nice interpretation in the
five-brane picture.  The $GL(2)$
transformation implied by \hecke\
rescales the spatial size of the string by $d$
units and the temporal size by $a$ units, in addition to shifting
the temporal direction into the spatial direction by $b$ units.  This
is equivalent to having
 $d$ wrapped M5-brane strings on top of each other and results in an
$SU(d)$ gauge symmetry as expected
from field theory analysis.   The factor of $a$ in the temporal  
direction
rescales the gauge coupling and the shift by $b$ picks out one of
the $d$ vacua by shifting the $\theta$ angle.

As an example, consider the case $n=p$, where $p$ is  prime.  In this
case the partition function takes the particularly simple form
\eqn\heckepart{
Z_p={1\over p^2} Z_1(p\tau)+{1\over p}\Big[Z_1\big({\tau\over p}\big)
+Z_1\big({\tau\over p}+{1\over p}\big)+...+Z_1\big({\tau \over p}+
{p-1\over p}\big)\Big] \ ,}
which correctly counts the BPS states for $p$-th wound heterotic
string.  This also agrees with the prediction made
in \CVEW\ for $SU(p)$ on $K3$.

\subsec{Comments about Coincident M5 branes and Extensions}

In the previous sub-section we have suggested that to
compute the partition function of $n$ M5 branes whose world-volume
is $K3\times T^2$, we simply use the result of one $M5$ brane
on $K3\times {\tilde T}^2$ where ${\tilde T}^2$ is n-times
wrapped over $T^2$.  As discussed above this is consistent
with the view that heterotic string emerges as an M5 brane wrapped
around $K3$.  However it would be nice to try and justify
this step directly from the viewpoint of M5 branes.  The potential
problem is that the $n$ M5 branes could interact non-trivially
and perhaps have non-trivial bound states, whereas the foregoing
result implies that the M5 branes do not interact, but simply
concatenate to produce an $n$-fold wrapping.  From the perspective
of the M5 branes, the main indication that the interactions will be
topologically trivial is that $n$ M5 branes on $K3$ can be  
holomorphically
separated:  the number of (supersymmetric) normal deformations
of $M5$ brane in the Calabi-Yau 3-fold is $h^{2,0}(K3)=1$.  Thus if
we use an element $u\in H^{2,0}$ we can holomorphically move branes
off one another.  It would be nice to make this argument rigorous
directly in the context of branes.  It is also natural to try to  
extend this
argument
to  manifolds with $h^{2,0} > 1$ as was done in \CVEW .  Once again,  
the
deformation in normal direction to the $M5$ branes can also be done
by choosing a $u\in h^{2,0}$ of the manifold. However, in general
the deformations will intersect each other along
some divisors $D_i$.  So we would expect the answer
for the partition function be similar to that of $K3$
modulo corrections coming from the M5 branes intersecting over $D_i$,
leading to new field theory subsectors.
This structure matches the results found in \CVEW\ for $SU(2)$,
and it is natural to expect to be able to justify the results
conjectured in \CVEW\ in connection with Montonen-Olive duality
using the M5-branes.

\subsec{Manifolds with $b_2^+=1$}

The partition functions of $N=4$ Yang-Mills on manifolds with
$b_2^+=1$ are not so easy to compute.  From the Yang-Mills point
of view it is on such manifolds that the gauge theory cannot be
deformed to an $N=1$ theory
by giving masses in a topologically invariant way.  From the
five-brane point of view, this is reflected by the fact that $n$
five-branes wrapped around such a manifold cannot be seperated in a
supersymmetric manner since there are no moduli for holomorphic
deformations. The  only such case that has been previously studied
is $SU(2)/Z_2$ on ${\bf P}^2$.  The partition function of
$SU(2)$ gauge theory on ${\bf P}^2$ is naively expected to be  a
holomorphic
function of the coupling, $\tau$, however it was found in
\CVEW\ that the partition function displays a holomorphic anomaly and
thus also depends on ${\overline \tau}$.

There are two choices of `t Hooft fluxes for ${\bf P}^2$.
We denote the partition function for each of these two choices by
$G_0=F_0/\eta^6$ (trivial flux) and $G_1=F_1/ \eta^6 $
(non-trivial flux).  It was found in \CVEW\ that
\eqn\PIIanom{\eqalign{{\overline \partial}_{\tau}F_0~=~ &{3\over
16\pi i}~
\tau_2^{-{3\over 2}}~ \sum_{n\in {\bf Z}} ~{\overline q}^{n^2} \ ,
\cr
{\overline \partial}_{\tau}F_1~=~ & {3\over 16\pi i}~
\tau_2^{-{3\over 2}}~\sum_{n\in {\bf Z}} ~{\overline q}^{(n+\half)^2}
\  .}}
Note that the power of $\tau_2$ in \PIIanom\ is fixed
by modular properties.  The existence of this anomaly
was interpreted in \CVEW\ as the contribution of reducible
$SU(2)$ connections in the form of $U(1)\subset SU(2)$ gauge
field configurations.   Although this
has not yet been verified, it seems to be a reasonable
conjecture.  Note that the theta-function sum on the
right hand side of the anomaly has the interpretation of $U(1)$
fluxes for $P^2$ (which has a single non-trivial 2-cycle).
{}From the viewpoint of two five-branes wrapped around ${\bf P}^2$,
this would correspond to the reduction $U(2)\rightarrow U(1)\times
U(1)$,
which comes from seperation of the two five-branes in spacetime.
Thus we interpret the existence of anomalies as a reflection of the
existence of non-trivial bound states at threshold, and these are
counted by
the
five-brane partition function.  This aspect will be discussed
more extensively later in this paper when we interpret the results
found for the BPS states of the E-string.

{}From the mathematical properties of instanton moduli space
on K\"ahler manifolds with $b_2^+=1$ it has become
clear that there are various chambers for the partition function
of topological Yang-Mills.  While this has been studied
mostly for $N=2$ theories (see in particular the
recent papers  \refs{\GMEW,\Nekr}), the $N=4$ case should parallel
the $N=2$ case.  Namely one  expects that the topological partition
function of $N=4$ Yang-Mills will depend on the choice of the
K\"ahler class, which is $b_2$-dimensional.  Thus, in general we
expect
that the $N=4$ partition function for $SU(n)/{\bf Z_n}$ Yang-Mills
on a K\"ahler manifold with $b_2^+=1$ depends not only on
${\overline \tau}$ but also on the K\"ahler class of the
manifold.

It is natural to conjecture the following form as a generalization of
the
holomorphic anomaly formula:  Let $\theta (q,{\overline
q})$  denote the partition function of $U(1)$ fluxes on a manifold
with $b_2^+=1$. This can be written in the form
$$\theta (q,{\overline q})~=~\sum_{P_L,P_R}~q^{\half P_L^2}~
{\overline q}^{ \half P_R^2}$$
where $P=(P_R,P_L)\in H_2(\cN,{\bf Z})$, and $P$ is split
to $P_R$ and $P_L$ by projecting  $H_2$ on to self-dual and
anti-self-dual parts using the K\"ahler form on $\cN$.  In
particular, the
topological self-intersection number is given by $P_R^2-P_L^2$.
The signature of the lattice is $(1,b_2-1)$, and
$\theta$ is a modular form of $\Gamma(2)$ (the lattice
is integral, but not necessarily even).  Once again
we decompose the $U(1)$ according to  fluxes based on how they are
paired with the
$SU(2)/Z_2$ 't Hooft fluxes.  Let $\alpha$ denote the choices of
inequivalent fluxes.  They are correlated with particular subspaces
$\Gamma_\alpha$ of $H_2(\cN,{\bf Z})$ (in a similar, but more
complicated manner than  $K3$).    Let
$$G_\alpha ~=~ F_\alpha/\eta^{2\chi}$$
denote the partition function of $SU(2)/Z_2$ with 't Hooft flux
$\alpha$, where $\chi$ is the Euler characteristic of the manifold.
Then there is a natural generalization for the holomorphic anomaly
of $P^2$, which we conjecture to be
\eqn\anomg{{\overline \partial}_{\tau}F_\alpha~=~{\rm const.}~
\tau_2^{-{3\over 2}} \sum_{P\in {\Gamma_{\alpha}}} ~
q^{P_L^2/4}~{\overline q}^{P_R^2/4} \ ,}
for some constant of proportionality
 which may depend on the K\"ahler moduli of the manifold.

\subsec{The partition function of $\half K3$}

As noted before, the interesting low energy limit
occurs for E-strings when they can become tensionless, and this
happens in the F-theory compactification when the Calabi-Yau
has a vanishing four cycle,
$\half K3= \cB_9$.  This manifold has  $b_2^+=1$ and is elliptic.
We will be interested in the properties of multi five-branes wrapped
around $\half K3$, and in particular, we will show how the
the BPS states of the E-string wrapped around a circle are enumerated
by the partition function of $U(n)$ Yang-Mills on $\half K3$.

The homology group, $H_2(\half K3, {\bf Z})$, is generated
the hyperplane section of ${\bf P}^2$, which we will
denote by $e_0$, and the nine blow-ups $e_j, j=1,\dots,9$.
(We will use the conventions of \Claude\ in which the
blow-ups are, more correctly, $-e_j$.)
The intersection form is thus $e_i\cdot e_j
= g_{ij}$, with $g_{ij}={\rm diag}(1,-1,-1,-1,-1,-1,-1,-1,-1,-1)$,
and thus  $H_2(\half K3,{\bf Z})$ is a hypercubic, Lorentzian,
self-dual, integral lattice with $(1,9)$ signature, and
we will denote it by $\Gamma^{9,1}$.

The lattice, $\Gamma^{9,1}$,  is isomorphic to the cubic
integral self-dual (but not even) $(1,1)$ lattice, $\Gamma^{1,1}$
plus the $E_8$ lattice $\Gamma^8$, \ie
$$H_2(\coeff{1}{2} K3,{\bf Z})~=~\Gamma^{1,1}\oplus \Gamma^8 \ .$$
To see this explicitly,
introduce the vectors $a_0$, $a_1$ and $b_i$, $i=1,\dots,8$,
defined by:
\eqn\basisvecs{\eqalign{ a_0& ~=~ 3e_0+\sum_{i=1}^8e_i \ ,
\qquad\qquad\qquad a_1 ~=~ -e_9 \ ,\cr b_i& ~=~ e_i-e_{i+1} \ ,
\ \  1\le i<8 \ ,\qquad\qquad b_8 ~=~ e_0+e_6+e_7+e_8}}
We will now show that this is an integral, unimodular change of
basis.  To see this, first note that $a_0$, $a_1$ and $b_i$
are integer linear combinations of the $e_i$.  Conversely, one
has $e_9=-a_1$ and
$e_8=3b_8+b_7+3b_6+5b_5+4b_4+3b_3+2b_2+b_1-a_0$.  The other $e_i$,
$i>0$, can then easily be obtained by adding integer
multiples of the $b_i$ to $e_8$.  Using these
expressions for $e_6$, $e_7$ and $e_8$ and subtracting them from
$b_8$,  we obtain an integer linear combination for $e_0$.
Thus the partition function of $\Gamma^{9,1}$ can be obtained
by summing over all integer linear combinations of the basis
vectors \basisvecs.

The new basis vectors satisfy $a_0\cdot a_0=1$, $a_1\cdot a_1=-1$
and $a_i\cdot b_j=0$.  We also have that $b_i\cdot b_i=-2$,
$b_i\cdot b_{i+1}=1$ for $i=1, \dots, 7$, and $b_5\cdot b_8=1$.
All other inner products are zero.  Hence the $b_i$ vectors are
the Dynkin vectors for the $E_8$ lattice and
$H_2(\half K3,{\bf Z})$ splits into $\Gamma^{1,1}\oplus\Gamma^8$.
Since $\Gamma^{9,1}$ is an odd, self-dual lattice, and $\Gamma^8$
is even self-dual, it follows that
the $\Gamma^{1,1}$ lattice is an odd, self-dual lattice.
For later use we note that the elliptic class $[E]$ and the base
of the elliptic fibration $[B]$ can be identified
with
$$[E]=3e_0+\sum_{i=1}^9e_i=a_0-a_1$$
$$[B]=-e_9=a_1$$
and that
\eqn\toni{[E]\cdot [B]=1\qquad [B]\cdot [B]=-1\qquad [E]\cdot [E]=0}

Once again we start by considering the partition function for a
single five-brane on $\half K3\times T^2$.  The degrees of freedom
on the string obtained by wrapping a five-brane around $\half K3$
involve twelve left-moving bosonic modes ($9=b_2-1$ left-moving
modes coming from the anti-symmetric field and the other three coming
from the three non-compact scalars) and four right-movers
(three non-compact coming from the three scalars and one compact
coming from the anti-symmetric field corresponding to $b_2^+=1$).
The right movers are accompanied by four fermions giving a $(0,4)$
supersymmetry.  If we compute the partition function of a single
five-brane we obtain
$$Z_1={\tau_2^{3/2} \over  V_3}~Tr \bigg[(-1)^F F_R^2 ~q^{L_0}~
{\overline q}^{\overline L_0} \bigg] ~=~
{{\theta (q,{\overline q})\over \eta(q)^{12}} \ ,}$$
where, as before, $\theta(q,\bar q)$ is a theta function over the
$H_2(\half K3,{\bf Z})$, which {\it does} depend on the choice of
the K\"ahler class of $\half K3$.
We have defined the partition function by dividing out the volume of
the three non-compact scalars which corresponds to the transverse
position of the five-dimensional string obtained by wrapping the
five-brane over $\half K3$.  As usual we have multiplied by the  
factor
$\tau_2^{3/2}$ to cancel that coming from the transverse momentum
integrations.  The modular weight of $Z_1$ is  $(9/2,1/2) + (-6,0) =
(-3/2,1/2)$ as it should  be.

The metric on $\half K3$ will induce a polarization on $\Gamma^{9,1}$
lattice, which is a decomposition into `left and right momenta' as
is familiar from Narain compactifications.
One of the K\"ahler moduli on $\half K3$ is $1/R$, the size of
the elliptic fiber.  The resulting polarization on
$H_2(\half K3, {\bf Z})$ respects the splitting
to $\Gamma^{1,1}\oplus \Gamma^8$.    Choosing a metric that
induces a polarization on $\Gamma^{1,1}$, we write $(p_L,p_R)=
\left({n+m\over2R}+{n-m\over2}R,{n+m\over2R}-{n-m\over2}R\right)$.
One can easily see that this is a basis for such an  integral, odd
self-dual lattice.  First, ${p_L}^2-{p_R}^2=n^2-m^2$, and so the  
lattice
is integral and odd.  One can easily show that if a vector  
$(p'_L,p'_R)$
is to have integer dot product with all vectors $(p_L,p_R)$, then
the former must also have the form $(p'_L,p'_R) =
\left({n'+m'\over2R}+{n'-m'\over2}R, {n'+m'\over2R}-{n'-m'
\over2}R\right)$.  Thus, the lattice is self-dual.

We therefore have:
\eqn\onfb{\theta (q,\overline q)~=~ \Big[ \sum_{n,m\in {\bf Z}}~
q^{\half ({n+m\over 2R}+{(n-m)R\over 2})^2} ~{\overline q}^{\half
({n+m\over 2R}+{(m-n)R\over 2})^2} \Big]~\theta_{E_8}(q)  \ ,}
where $\theta_{E_8}$ denotes the theta function for the $E_8$
lattice.  Note that the contribution of $\Gamma^{1,1}$ to the
$\theta$-function does {\it not} have the usual Narain form for a
string on a circle.  The reason for the difference is that the Narain
lattice is even and self-dual, while $\Gamma^{1,1}$ is only
odd self-dual.

As was noted in section 2, we will be interested in the
F-theory limit, where the size of the elliptic fiber shrinks to zero,
$1/R\to 0$.  This is the limit that corresponds to the
six-dimensional tensionless string.
One should note that $U(1)$ fluxes in the left-over $E_8$ lattice can
still be non-zero and will correspond
to  K\"ahler classes of the ${1\over 2} K3$ for infinitesimal,
but non-zero size of the elliptic fiber.
In the limit $R\rightarrow \infty$
the sum over $n,m$ reduces to a single sum over the null
direction, $n=m$.   This sum can be computed and gives
$R\over \sqrt{2\tau_2}$. Thus one finds that for $R\rightarrow
\infty$,
$$Z_1={R\over {\sqrt{2\tau_2}}}{\theta_{E_8}(q)\over
\eta(q)^{12}\ }. $$
Note that if we view $R$ as the radius of a circle that takes us
from F-theory in six dimensions down to five dimensions,
then the extra factor of $R$ is the volume factor that comes from
string compactification on this extra circle. Moreover, the factor
of $1/\sqrt{\tau_2}$ is manifestly coming from the continuum
momentum integration in the decompactified direction.
Now recall that we have adopted the convention of defining
the $F$-theory partition function, \zpartf, with a prefactor of
$\tau_2^2/V_4$ while the M theory partition function has a prefactor
of $\tau_2^{3/2}/V_3$.  The factor $R / \sqrt{\tau_2}$ represents
precisely this difference in these prefactors.
Thus dropping the factor of $R / \sqrt{\tau_2}$ should
yield the partition function of a string in six dimensions.
Doing this, we then view
\eqn\ZoneBnine{Z_1~=~ {\theta_{E_8}(q)\over \eta(q)^{12}} \ .}
as the partition function for a six dimensional E-string. Note that
this is a holomorphic modular form of weight $(-2,0)$ as we had
anticipated earlier.

We would now like to generalize this to $n$ five-branes
wrapped around $\half K3$, and we start by considering $n=2$.
As already mentioned, since we are considering a K\"ahler manifold
with $b_2^+=1$ there is no conjectured answer coming from the $U(n)$
Yang-Mills theory.  Based on the foregoing, we should expect a
holomorphic anomaly, as well as some dependence on the choices of
K\"ahler classes.
However, since we have gone to a particular degenerate limit of the
K\"ahler class, we have effectively
frozen out all the K\"ahler class dependence
which is responsible for the existence of various chambers, and so we
should only expect a holomorphic anomaly.

The $U(2)$ theory on $\half K3$ requires a rescaling $\tau
\rightarrow \tau/2$ as on $K3$.  The partition function will
decompose, as discussed
above,  in terms of $SU(2)\times U(1)/Z_2$ where the 't Hooft fluxes
for $SU(2)/ Z_2$ are correlated with the $U(1)$
fluxes.  In the limit $R\to\infty$, all  $U(1)$ fluxes reside
in a particular null direction ($m=n$) on the
$\Gamma^{1,1}$ lattice.
Following the decomposition \onfb\ of the $\Gamma^{9,1}$
lattice, we write the `t Hooft fluxes  $\alpha = (\lambda,i,j)$
where $\lambda$ is a vector on the $E_8$ lattice, and $i,j = 0,1$
denote the even/odd terms along the space-like and time-like
directions of $\Gamma^{1,1}$.
Since we are restricting to $m=n$ on $\Gamma^{1,1}$, we are only
interested in the $SU(2)/Z_2$  partition functions $Z_{\lambda  
,j,j}$.
For a given set of $U(1)$ fluxes, both even and odd values of $j$
give the same  result in the limit $R\rightarrow \infty$,
since one is simply taking the continuum limit of the sum.
Therefore, we effectively get the sum of `t Hooft fluxes in this
direction, and so we define:
\eqn\combi{\hat Z_{\lambda}={\half}(Z_{\lambda,0,0}+
Z_{\lambda,1,1})\ .}
Let $\beta = (\lambda',k,\ell)$, then one has $(\lambda,j,j) \cdot
\beta = \lambda \cdot \lambda' + j(k -\ell)$
(remember this is a Lorentzian inner product).  Using the modular
transformation \shar\ on $\hat Z_{\lambda}$ we therefore see that
$Z_\beta$ has an overall coefficient of $\sum_j exp(2 \pi i
(\lambda \cdot \lambda' + j(k -\ell))/2)$, which projects onto a
sum of those $Z_\beta$ with $k = \ell$.
It follows that $\hat Z_\lambda$ maintains its form
under $SL(2,Z)$ transformations, and indeed transforms with
phases that depend only on the $E_8$ label $\lambda$.
Henceforth we will drop the hats on $\hat Z_\lambda$ with the
understanding that $Z_\lambda$ means \combi. (Similarly, for
$SU(n)$ we consider $\hat Z_{\lambda}={1\over n}\sum_m
Z_{\lambda,m,m}$.)

After  the rescaling by the volume factor, the partition function
of two five-branes on $\half K3\times T^2$, including $U(1)$ fluxes,
becomes
$$Z_2=\sum {Z_\lambda} ~\theta_{E_8,\lambda}(q^{\half})$$
where $\lambda$ denotes the inequivalent $Z_2$ fluxes on the
the $E_8$ part of $H_2$, and $\theta_{E_8,\lambda}$ denotes the theta
function for the corresponding $U(1)$ fluxes.
There are three inequivalent choices for $\lambda$: $0$, even and  
odd,
exactly as there were with $K3$. One can thus readily
write down the corresponding  $\theta_{E_8,\lambda}$.
The question remains as to how to determine $ Z_{\lambda}$.

What properties do we expect of $Z_2$?  The list includes:
i) $Z_2$ is the
partition function of $U(2)$ and  so, using Montonen-Olive
duality, we expect it to be a modular
form of $SL(2,{\bf Z})$.
ii) The total modular weight of $Z_2$ should be the sum of the weight
$(-\chi/2 ,0)=(-6,0)$
which is the $SU(2)$ contribution and the weight
$(4,0)$ which is the
contribution of the $U(1)$ fluxes.  Hence, the total weight is
$(-2,0)$,
which is consistent with the partition function in \partfs.
iii) From \CVEW\ the smallest power in its $q$ expansion
is expected to be $q^{-2\chi/24}=q^{-1}$.  iv) The instanton
expansion should have a ``gap'' since the
first non-vanishing instanton number is $2$ \ref\gapins{T. Pantev,
private communication.}
v)  $Z_\lambda$ should  have a holomorphic anomaly in the form
\eqn\holanom{
{\overline \partial}_\tau Z_\lambda ~=~{i \over 2 \pi \tau_2^2 }~
C_\lambda~{\theta_{E_8,\lambda}(q^{\half})\over \eta^{24}} \ ,}
where the extra power of $\tau_2^{-\half}$ as compared to \anomg\
comes from the theta function in the limit  $R\to\infty$.
Note that the modular weight on both sides of \holanom\ is $(-6,2)$.
These conditions go a long way towards fixing
$Z_2$.  In fact if we knew the constants $C_\lambda$, together
with modularity properties and the gap for $Z_2$ it would fix it
completely.  In the next section we will derive
the constants, and compute ${ Z}_\lambda$ explicitly.   It turns
out that the holomorphic anomaly is most conveniently stated
for $Z_2$, and we find that
\eqn\holanomII{
{\overline \partial}_\tau Z_2 ~=~{{\rm const.}\over \tau_2^2}~
Z_1^2}
This has the interpretation of two five-branes wrapped around $\half
K3$ forming bound states at threshold.  We will
discuss this later in the paper. Also note that the modular weight
on both sides of \holanomII\ is $(-2,2)$.

What should we expect for $Z_n$, the partition function for $n$
five-branes wrapped around $\half K3$?  First, in taking the
$F$-theory limit to get the six-dimensional partition function,
there will now be $n$ terms in the sum \combi.  The argument above
easily generalizes to show that these sums close under the
modular transformation \shar.  These coefficient functions
are then to be combined with appropriate $\theta$-functions
of the suitably scaled $E_8$ lattice, and result is expected to be a
modular form of $SL(2,{\bf Z})$ of weight $(-2,0)$, whose holomorphic  
part
(\ie fixing $\tau$ and sending ${\overline \tau}\rightarrow \infty$)
starts at $q^{-n/2}$.  We also expect there to be a gap of $n$
units in the $q$-expansion because the first non-zero
instanton number is $n$.  In other words, the next non-zero term will
be the $q^{n/2}$ term.  Furthermore, we expect
there to be a generalization of the holomorphic anomaly corresponding
to reducible connections.   For $U(n)$ the natural reducible
connections  correspond to $U(n)\rightarrow U(k)\times U(n-k)$,
so we expect ${\overline \partial}_\tau Z_n$ to involve
a weighted sum over $Z_k\cdot Z_{n-k}$.  As we will
explicitly show, this is correct and the equation is
\eqn\holanomgen{{\overline \partial}_\tau Z_n ~=~ {{\rm const.}\over
\tau_2^2}~
\sum~ k (n-k) ~ Z_kZ_{n-k}}
{}From the five-brane point of view this can be  interpreted as
the possible ways $k$ bound  strings can bind with $n-k$ bound
strings in a pairwise manner to form $n$ bound strings.

\newsec{Counting BPS states of the Compactified $E$-string}

In this section we summarize the basic approach
in counting the BPS states of the E-string, continuing
the discussion in section 2, and sketch some of the
properties of what we shall find.  In this section
we try to avoid the technical aspects of this computation,
postponing them to the subsequent section.

\subsec{Computing the pre-potential}

We consider here the $E$-string compactified on a circle
to five dimensions and  count the BPS states with a given
winding number and momentum.  Comparing this to
the structure and form of the Yang-Mills partition functions,
one can read off the ${\hat Z}_\lambda$.

The basic approach that we will use for counting the BPS states
was first employed in \KMV, and further developed in
\refs{\LMW,\MNWa,\MNWb}.  The basic idea is that if we consider
type IIA theory on a Calabi-Yau threefold, the pre-potential
of the four-dimensional $N=2$ theory only receives world-sheet
instanton corrections.  Indeed the pre-potential can be
represented as a sum over holomoprhic curves, and can thus
be computed using mirror symmetry techniques.
One actually needs a local version of the mirror symmetry
to do this counting because the relevant part of the Calabi-Yau
geometry, in the limit of turning off the gravity, is a non-compact
manifold corresponding to a Calabi-Yau singularity and its immediate
neighborhood.  Such a non-compact specialization was used in  
\refs{\KMV,\LMW}\
while more general  aspects of local mirror symmetry and geometric
engineering were  developed in \refs{\KKV,\KatzMV,\LeVa}, where it
was used to solve the Coulomb branch of $N=2$ theories in four
dimensions.

Let $t_i$ denote the complexified K\"ahler moduli of
Calabi-Yau threefold corresponding to the $i$-th 2-cycle
 and $N_{[n_i]}$ denote the number of holomorphic
curves wrapped around the $i$-th cycle $n_i$ times, then
the  third derivative of the pre-potential is given by
\eqn\yeki{\partial^3_{ijk}{\cal F}(t_i)=\sum_{[n_i]} n_in_jn_k
N_{[n_i]}~{\prod_i q_i^{n_i}\over 1-\prod_i q_i^{n_i}}}
where $q_i={\rm exp}(-t_i)$.  Applying this to $\half K3$ embedded
in a Calabi-Yau manifold, and
given the relation between holomorphic curves and the BPS states
of the E-strings, discussed in section 2, this will
amount to the computation of the BPS states of the multiply
wrapped E-string around a circle. This expression
also includes the contribution of multi-wound states (in the form
of the denominator above) and is natural to view this function
itself as the partition function of BPS states (from which the
primitively wound BPS string states can be obtained).  Note also
that \yeki\  fixes ${\cal F}$ up to a quadratic function of
$t_i$, and the latter can be viewed as the classical contribution
to the pre-potential.

Of the ten K\"ahler moduli of $\half K3$, the two that
index the winding state and momentum state of the compactified
string can be thought of as the moduli of
the base and fiber respectively of the elliptic fibration.
There are in addition eight remaining K\"ahler moduli.  The
intersection
matrix is an integral inner product in a ten dimensional lattice
with signature $(1,9)$ and it includes a sublattice on which
the intersection matrix is that of
the Cartan Matrix of $E_8$.  For future reference, we will
label these moduli by $\phi,\tau$ and $m_i, i =1, \dots,8$ where
$\phi$ and $\tau$ correspond to the base and fiber moduli and
the $m_i$ label the $E_8$ moduli.  In the next section
we will review the construction of mirror map for ${1\over 2} K3$
and also present a simple new derivation for it
based on $R\rightarrow 1/R$ duality of tori, very
much in the spirit of examples studied in
\VWtor.  The basic aspect of the new derivation is that
${1\over 2}K3$ arises in a self-mirror Calabi-Yau 3-fold
and thus the K\"ahler moduli of it are mirror to its own complex
moduli.  In other words, roughly speaking, studying the {\it complex
moduli}
of ${1\over 2} K3$ amounts to obtaining the pre-potential.  More
precisely ${1\over 2} K3$ is an elliptic manifold and the
corresponding Seiberg-Witten
curve for the $N=2$ theory is simply obtained
from the complex structure of ${1\over 2}K3$ where the coordinate of
the base of ${1\over 2}K3$ is promoted to a moduli parameter.

Recasting the mirror map in terms of periods of a torus
enables one to easily study the modular properties of
the partition functions as a function of $\tau$, and by
making asymptotic expansions for large $Im(\phi)$, one can explicitly
obtain a series, indexed by the winding number, $n$, of the
$E$-string, whose coefficients are functions of the moduli, $m_i$
and $\tau$.  Let ${\cal F}$
be the pre-potential  of the theory, and suppose that it has an
expansion of the form:
\eqn\prepotexp{{\cal F} ~=~ {\cal F}_{\rm classical} ~+~
\sum_{n= 1}^\infty ~q^{n/2}~Z_n(m_i;\tau)~
e^{2 \pi i n \phi} \ ,}
where $q=e^{2 \pi i \tau}$, and the function
${\cal F}_{\rm classical}$ is a cubic in the moduli,
whose third derivative gives the classical intersection form.
The functions $Z_n(m_i;\tau)$ count the rational curves in the
$\half K3$ which wrap $n$-times around
the base. For $m_i = 0$, the functions $Z_n(0;\tau)$ are
(almost) modular forms\foot{To arrange this one has to use the
somewhat unusual normalization factor of $q^{n/2}$ in \prepotexp.}
 of weight $-2$: That is, they can be written in the form:
\eqn\fnform{Z_n(0;\tau) ~=~ {1 \over \eta^{12 n}} ~ p_n (E_2(\tau),
E_4(\tau),E_6(\tau)) \ ,}
where $p_n$ is some (quasi-) homogeneous polynomial of weight $6n-2$,
and $E_{2m}$ are the Eisenstein modular forms of weight $2m$.  The
function $E_2(\tau)$ is holomorphic, but transforms with the usual
modular anomaly.  This modular anomaly, and the fact
that $\eta^{12}$ changes sign under $\tau \to \tau+1$ define
the sense in which we mean that ${\cal F}$ is
{\it almost} a modular form.  One can remove the modular anomaly
from $E_2$ by adding to it a suitable multiple of $1/Im(\tau)$
to yield the function $\widehat E_2$.  In this way we can ``ignore''
the modular anomaly, but at the cost of introducing a rather
mild anti-holomorphicity.  We will denote the corresponding
modified form of $Z_n$ in this section by $\hat Z_n$.  In the
subsequent
section we revert back to the $Z_n$ notation
for the partition function, including
the $\tau_2$ piece.

\subsec{The recursion relation}

It was shown in \refs{\MNWb,\MNWc}
that the $E_2$ content of $Z_n$ is, in fact, entirely determined
by the recurrence relation:
\eqn\recursion{ {\partial Z_n \over \partial E_2} ~=~ {1\over24}~
\sum_{m=1}^{n-1}~ m(n-m)~Z_mZ_{n-m} \ .}
The corresponding equation for $\hat Z_n$ is
\eqn\hatrecursion{ {\partial \hat Z_n \over \partial \bar
\tau} ~=~ - {i \over 16 \pi}~{1 \over (Im(\tau))^2}~
\sum_{m=1}^{n-1}~ m(n-m)~\hat Z_m \hat Z_{n-m} \ .}
Note also that if we kept the $\phi$ dependence in
the partition funcation and consider $Z=\sum_n\hat Z_n
{\rm exp}(2\pi i n\phi)$ then the foregoing equation takes the form
$${\partial Z\over \partial \overline \tau}={i\over 64 \pi^3
(Im\tau)^2}{\partial Z\over \partial \phi}
{\partial Z\over \partial \phi}$$
It is clear from \MNWc\ that such a recurrence relation is a
very general feature of an effective action of a theory with a
parameter given by the modulus of a torus.  In particular, the
proof of \MNWc\ is easily generalized to incorporate the
 K\"ahler moduli, $m_i$.  The non-zero parameters, $m_i$, mean
that the $Z_n$ transform like group characters:
\eqn\modprop{\eqalign{Z_n(m_i,\tau +1) ~=~ (-1)^n ~Z_n(m_i,\tau) \ ,
\cr
Z_n\Big({m_i\over \tau},-{1\over \tau}\Big ) ~=~ \tau^{-2} ~
e^{{i  \pi \over \tau} \sum_j n\, m_j^2}~
Z_n(m_i,\tau) \ ,}}
where we have ignored all the modular anomalies coming from $E_2$
(which can be done by promoting it to $\widehat E_2$).

{}From the Yang-Mills point of view, \hatrecursion, is the
$SU(n)$ generalization of the holomorphic anomaly equation \anomg\
anticipated in \CVEW , which was conjectured to arise
from reducible connections.  Given that the BPS
counting of E-strings is related to the counting of holomorphic
curves, it is also natural to try interpreting the foregoing anomaly
equation from this viewpoint.  Indeed there turns out to
be a simple interpretation from this angle:  Recall
that $Z_m$ denotes the number of holomorphic curves
which wrap $m$ times around the base of $\half K3$ and
an arbitrary number of times around the elliptic fiber (which
gives the $\tau$ dependence of $Z_m$).  Given
the T-duality of the fiber torus, we would expect
$Z_m$ to be a modular form of $\tau$.  However, there
is a natural way in which we can obtain bound states of holomorphic
curves corresponding to $Z_m$ and $Z_{n-m}$ and that is by joining
them along a fiber at two different points along the fiber;
due to translation invariance of the torus only the relative
position of these two points is relevant.   The resulting
holomorphic curve will wind around the base of $\half K3$,
$m+(n-m)=n$ times.  Note that we can attach the torus
in $m$ ways to the first curve, because there are $m$ leaves
of the first holomorphic curve around the sphere, and $n-m$ ways
to the second one.  So if we consider how we may remove a
torus from $Z_n$ so as to obtain $Z_m$ and $Z_{n-m}$, we expect
to have an equation of the form
\eqn\remtorus{{\partial Z_n \over \partial F_1}~=~
\sum_m m(n-m)Z_mZ_{n-m} \ ,}
where $F_1$ is the partition function of the torus.  That is, $F_1$  
is
the number of ways we can wrap the torus around the fiber torus
(modulo translation and the choice of a point on it).  This function
is known to be \BCOV\  $E_2/24$ (up to the holomorphic anomaly term).
Using this, we see that \remtorus\ is precisely the recusrion  
relation
\recursion .  Thus the holomorphic
anomaly here has exactly the same origin as that observed in \BCOV
\foot{We should recall that we have been using genus zero
curves to count the holomorpic curves in the Calabi-Yau.  These
do seem to receive contributions which also look like higher
genus curves.  The reason that this is not surprising is
that the genus zero partition function gives the pre-potential
of the $N=2$ theory and that has the information about all the
{\it charged} BPS states.  This is also presumably why the  
configuration
involving connecting $Z_i$'s around a ring with $T^2$ fibers
is not contributing to the BPS states.  These would look
like genuinely higher genus curves and will only modify gravitational
corrections.}.

\subsec{The ``gap'' }

The other key property of the $Z_n$ is its ``gap:''  That is,
its $q$-expansion ($q = e^{2 \pi i \tau}$) has the form:
\eqn\gapcond{Z_n(m_i; \tau) ~=~ {1 \over n^3}~q^{-{n \over 2}} ~+~
O(q^{+{n \over 2}}) \ .}
One way of understanding the ``gap'' comes from recalling how
${\cal F}$ counts rational curves \refs{\KMV\LMW\MNWa\MNWb}.
This is the equation given in \yeki\ which we now
specialize to the case at hand.  The quantum part of ${\cal F}$,
${\cal F}_{quantum} ~=~  {\cal F} - {\cal F}_{\rm classical}$, has
an expansion of the form:
\eqn\prepot{{\cal F}_{quantum} ~=~ \sum_{j,k,\ell_1,\dots,
\ell_8 =0}^\infty\, N_{j,k,\ell_1,\dots, \ell_8} ~
\sum_{r = 1}^\infty ~ {1 \over r^3}~
e^{2\pi i r ( j \phi + k \tau + \vec \ell \cdot \vec m) }\ , }
where $N_{j,k,\ell_1,\dots, \ell_8}$ is the number of irreducible
rational curves with winding numbers $j,k,\ell_1,\dots, \ell_8$
on the cycles with K\"ahler moduli $\phi,\tau, m_i$.

To see the gap property, one has to note that the intersection
of two distinct holomorphic curves is always positive
(since they are holomorphic the intersection points
all have the same relative orientation).  So let us
consider the holomorphic curve corresponding to $j[B]+k[E]$ where
$[B]$ denotes the class of the base and $[E]$ denotes the elliptic
fiber.  Consider another holomorphic curve which is $[B]$ itself.
As long as $k\not=0$ these two curves are distinct and we should thus
have
$$(j[B]+k[E])\cdot [B]\geq 0$$
Now we use the fact \toni\ that $[B]\cdot [B]=-1$ and $[B]\cdot  
[E]=1$ and
learn that
$$k-j\geq 0 \qquad if \quad k\not=0$$
This establishes the existence of the gap.

One can calculate the $Z_n$ directly by computing the expansion
(at large $Im(\phi)$) of the period integrals
of the appropriate torus.  This will be done in the next
section to obtain $Z_1(m_i,\tau)$.  We have
also derived  $Z_2(m_i,\tau)$ in the same manner,
but such direct calculations
rapidly becomes unwieldy.  As was seen in \refs{\MNWb,\MNWc},
the most powerful method for explicitly computing the $Z_n$ is
to use the recurrence relation \recursion\ and the gap condition
\gapcond.  That is, once one has $Z_1$, one can make an
Ansatz for $Z_n$ that is consistent with the modular properties
\modprop\ and the $E_8$ Weyl invariance, and then fix the
coefficients
using \recursion\ and \modprop.  This generically yields a highly
overdetermined system of equations which give a unique result for
$Z_n$. We illustrate this in some detail for $Z_n$ in
section 6 for small values of $n$.

\newsec{Deriving the Wound E-String BPS States}

In this section we spell out in a bit more detail the technique
used in counting the BPS states of multiply wound E-strings.
Note that as discussed before this corresponds to partition function
of $N=4$ topological $U(n)$ Yang-Mills on $\half K3$ where
$n$ corresponds to winding number of E-string around the circle.
In the next section we apply the method to explicitly
write down the BPS state for E-strings with low winding
numbers.

The appropriate mirror maps were constructed in \KMV\ for
subsets of the moduli.  Non-compact Calabi-Yau
manifolds that modelled the non-critical string were constructed
in \LMW\ and the construction of mirror map was reduced to
computing periods of a Seiberg-Witten torus.  The models of
\LMW\ gave a subset of the BPS spectrum
(those states with winding number equal to the momentum) and
only involving the modulus $\phi$. In \MNWa\ the Calabi-Yau
models of \LMW\ were extended to incorporate all the BPS states
and both the $\tau$ and $\phi$ moduli.  Moreover, after
reducing the periods of the Calabi-Yau to periods of
various differentials on a torus, it was shown in
\refs{\MNWa,\MNWb} how to incorporate the complete set of
ten  K\"ahler moduli.  Thus, in \refs{\MNWa,\MNWb} the construction
of the complete mirror map, involving all moduli, was reduced
to computing periods of a Seiberg-Witten differential on
a torus.  This torus had also been constructed by other
authors \refs{\GMS} in order to get effective actions of gauge
theories arising out of the $E$-string.

It was shown in \refs{\LMW,\GMS,\MNWa,\MNWb,\MNWc} that the  
pre-potential
of the $E$-string could be derived from a Seiberg-Witten
curve of the form:
\eqn\SWcurve{y^2~=~x^3~-~f(u,m_i,\tau)x~+~g(u,m_i,\tau) \ ,}
where $u$ is a parameter that essentially sets the overall size of
the
del Pezzo surface (an overall scale has been set to one).
This curve in fact describes the del-Pezzo surface $B_9$.
The Seiberg-Witten curve depends on $u$, the eight
mass parameters $m_i$ and a parameter $\tau$, which is the K\"ahler
modulus of the elliptic fiber.  The masses are inversely related to
the K\"ahler moduli for the blow-up moduli of the $B_9$.
That is, taking masses to infinity is equivalent
to blowing down some of these points.  We first present a simple
new derivation of the above local mirror, and then discuss how
one computes the instanton expansions from it.

\subsec{The Mirror Map for ${1\over 2} K3$}

In order to present a simple derivation of these results we need
to find a Calabi-Yau threefold for which ${1\over 2}K3$ is embedded.
Consider a Calabi-Yau threefold which is a double elliptic
fibration over a sphere:
$$y_1^2=x_1^3-f_1(z)x_1+g_1(z)$$
\eqn\namit{y_2^2=x_2^3-f_2(z)x_2+g_2(z)}
where $(y_i,x_i)$, subject to the above equations,
define the two tori and $z$ denotes the coordinate
on the sphere.  Moreover $f_i$ are functions of degree
4 and $g_i$ are functions of degree 6 in $z$.  This is a Calabi-Yau
threefold with Hodge numbers $h^{1,1}=h^{1,2}=19$.  There is a
$\half K3$ embedded in the Calabi-Yau of \namit, since each of
the equations in \namit\ describes a $\half K3$ surface in three
complex dimensions.
equation.
The local model, i.e.
 the isolated ${1\over 2} K3$,
is obtained by enlarging the second torus, in other words
by sending the K\"ahler class of the second torus to infinity.
For a particular
choice of $f_i,g_i$ this manifold can be obtained as a $Z_2\times
Z_2$
orbifold of $T^2\times T^2\times T^2$:  Let $\zeta_1,\zeta_2,
\zeta_3$ denote the coordinates of the three tori.
Consider the generators of the two $Z_2$ actions,
$$\alpha:  \zeta_1\rightarrow \zeta_1+{1\over 2},\quad \zeta_{2,3}
\rightarrow -\zeta_{2,3} \ ;$$
$$\beta:  \zeta_2\rightarrow \zeta_2+{1\over 2},\quad \zeta_{1,3}
\rightarrow -\zeta_{1,3} \ ;$$
where for definiteness we have taken the $d\zeta_i$
to have periods 1 and $\tau_i$.
It is easy to check using standard orbifold techniques,
that this orbifold has hodge numbers $h^{1,1}=h^{1,2}=19$.
Moreover, it corresponds
to a double elliptic fibration over a base ${\bf P}^1$.  Namely,
we can identify $\zeta_1$ with the ($x_1,y_1$) torus,
$\zeta_2$ with the ($x_2,y_2$) torus and $\zeta_3$ with $z$
(which parameterizes ${\bf P}^1$).  To see this note that
modding out by $\alpha$ leads to an elliptic
($\zeta_1$) fibration over the sphere ($\zeta_3$ modded by $Z_2$)
and modding out by $\beta$ leads to another
elliptic fibration ($\zeta_2$) over ${\bf P}^1$ (again
parameterized by $\zeta_3$ modulo the $Z_2$ action).
The double elliptic fibration, together
with the equivalence of hodge numbers, identifies the orbifold
with the Calabi-Yau in \namit.  Now we apply mirror symmetry
 as in \VWtor, by $T$ dualizing  one circle
from each of the three tori.
As explained in \VWtor\ this leads to the mirror manifold
if in addition one includes a non-trivial $Z_2$ discrete
torsion $\epsilon(\alpha , \beta )=-1$.  Since there are
no mutual fixed points for $\alpha$, $\beta$ or $\alpha \beta$,
this will not change the geometry of the massless modes.  Hence, we
obtain the same manifold \namit\ , at least as far as the
pre-potential is concerned,  but now with the complex moduli
playing the role of what were the K\"ahler moduli.   Note in
particular that the K\"ahler and complex moduli
of each of the two tori $\zeta_{1,2}$ are exchanged.
The description of the manifold
in terms of \namit\ is now quite useful because the K\"ahler moduli  
of
the original Calabi-Yau now appear explicitly in the coefficients
of $f_i$ and $g_i$.

On the mirror side the interesting object is the
holomorphic 3-form, the periods of which encode
the holomorphic curves (see for example \STY).
In this case, the holomorphic 3-form $\Omega$ is given by
\eqn\Omegaeq{
\Omega ={dx_1\over y_1}{dx_2\over y_2}dz}
which is regular for large $z$.
As noted before we are interested in a large
K\"ahler class for the second torus, which from the
mirror perspective, corresponds to a degenerate complex structure
for this same torus.  Suppose the second torus
becomes degenerate at $z=u$.  Locally this means that the torus is
parameterized as
\eqn\localmodel{
y^2-x^2=\tilde x\tilde y=(z-u)}
where $\tilde x$ and $\tilde y$ are the re-defined coordinates
on the second torus, emphasizing its degenerate
structure. At $z=u$ the torus is strictly
degenerate since one of its cycles has shrunk to zero.
Using the local model in \localmodel\  when $z\not=u $
means that the complex structure of the second torus remains  
degenerate
away from $z=u$.

We now investigate the relevant periods.  The holomorphic three-form,
$\Omega$, takes the form
\eqn\localOmega{
{dx_1\over y_1}dz {d{\tilde x}\over \tilde x}}
in the local limit.
There are two types of 3-cycles.  One type corresponds
to taking a cycle of the form $C_2\times S^1$ where $C_2$ is a two
cycle supported on the mirror $\half K3$ given by the variables
$(x_1,y_1,z)$, and $S^1$ is a cycle of the degenerate torus.  The  
other
type of three cycle is an $S^3$, which can be thought of as
$S^1\times S^1$ over an interval, where on one end of the interval
the first circle shrinks to zero and on the other end the second
circle shrinks to zero.  We choose one of these two circles to be
the small cycle of the second torus which is zero at  $z=u$, and we  
choose
the second cycle to be the $a$ or the $b$ cycle of the first torus.
(This construction and computation of the period integrals is very
similar to that of \LMW.)

Since we have only one non-degenerate torus we
remove the subscripts from $x_1,y_1$ and $f_1,g_1$
 and denote them by $x,y$ and $f,g$.
We first consider cycles of the form  $C_2\times S^1$.
Integrating $\Omega$ over the small cycle of this torus
gives  a constant, multiplied by the left-over 2-form
$$\int_{1-cycle}\Omega= \Omega' ={dx\over y}dz \ .$$
This 2-form is then integrated over 2-cycles of ${1\over 2}K3$.
%
%
Even though there are ten 2-cycles on this space,
$\Omega'$ vanishes on two of them, namely the 2-cycles
 corresponding to the fiber
torus and the base sphere.  So this type of cycle gives eight
non-zero parameters.  This
is  the kind of structure we expect from Seiberg-Witten
geometry for masses \SW,  so we identify these periods
with eight mass parameters $m_i$ (These can be
viewed as the dual to the Cartan subalgebra of $E_8$).
To get cycles of the second type we have to integrate
$\Omega'$ from $z=u$ to one of the points where the first torus
degenerates.  Call that point $P$.
We thus have
$$\phi=\int_P^u dz\int_{a}{dx\over y}.$$
Which point $P$ we choose is irrelevant because the difference
between two choices
amounts to an integral over a 2-cycle of the first type, which
means shifting $\phi$ by integral multiples of $m_i$.
This is also familiar  from
Seiberg-Witten geometry.  We can similarly define
$\phi_D$ by integrating over a $b$ cycle of the torus and choosing
a point where the $b$ cycle vanishes.  To make contact with the
usual notation of Seiberg-Witten, we re-define the dummy variable
$z$ by $u$ and summarize the periods by
$$m_i=\int_{C_i} du {dx\over y}$$
$$\phi=\int_{P}^u du\int_{a}{dx\over y}$$
where
$$y^2=x^3-f(u)x+g(u)$$
Note also that the $\tau$ which is mirror to the
K\"ahler class of the original $\half K3$ is now mapped
to the complex structure of the torus defined above.  The
functions $f$ and $g$ are now implicitly a function of
$f(u,m_i,\tau)$, $g(u,m_i,\tau)$.

\subsec{Computing the periods}

In order to count BPS states, we need to compute the above
periods for
the Seiberg-Witten curve.  We also want to group the BPS states into
$E_8$ Weyl orbits.  This is accomplished by turning on the mass
parameters and expressing the instanton expansion in terms of
$E_8$ characters. This was done for the special case
$\tau\to i\infty$ in \MNWa.  Now we want to
consider the more general case for finite $\tau$.

There is one major issue that needs to be addressed. For general
$m_i$ and $\tau$ the functions $f(u,m_i,\tau)$ and $g(u,m_i,\tau)$
are not known explicitly.  In principle, they can be determined from
the curve in \GMS, but in order to do so, one needs to first perform
a non-trivial $SL(3)$ transformation on that curve.  In \MNWa\
it was shown how to compute these functions in the
$Im (\tau) \to \infty$ limit explicitly from the instanton expansion.
Essentially one assumes that the instanton expansion
will give $E_8$ characters and one then works backwards to find the
curve.  The key point is that one does not need really need all of
the masses to be non-zero in order to find the characters, if one
assumes that the characters are actually $E_8$ characters.
At least for relatively
low instanton numbers two non-zero mass parameters are enough.

The functions $f$ and $g$, for two non-zero masses,
were derived in \MNWa.  In this case the Seiberg-Witten curve has the
relatively simple form
\eqn\SWIIm{y^2~=~x^3~+~\c u^2x^2~+~k^2u^4x-~2u(u^2~+~{s_+}^2x)
(u^2~+~{s_-}^2x) \ ,}
where $\c=1+k^2$,
\eqn\defk{ k \  = \ {\vartheta_2^2(0) \over \vartheta_3^2
(0) }\ , \qquad\qquad s_\pm=sn(m_\pm) \  = \
{\vartheta_3(0) \over \vartheta_2(0) }
{\vartheta_1(m_\pm) \over \vartheta_4(m_\pm) \ , }}
and where $m_\pm=(m_1\pm m_2)/2$.  We can then shift $x$ in \SWIIm\
to obtain the canonical form in \SWcurve.  If we shift
$x\to x-{1\over3}\c u^2+{2\over3}{s_+}^2{s_-}^2u$ and rescale $u$ to
$u\to u\vartheta_3^{12}$, one can recast \SWIIm\ into the form of
\SWcurve.  The actual expressions for $f(u,m_1,m_2,\tau)$ and
$g(u,m_1,m_2,\tau)$ are given in  appendix A.

An elliptic curve with modulus $\ttau$, up to a rescaling has the
canonical form
\eqn\eccan{
y^2~=~x^3~-{1 \over 3 \om^4 } \tE_4x~+~{2\over27\om^6} \tE_6 }
where $\tE_n=E_n(\ttau)$ is the Eisenstein series of modular weight
$n$
 and $\om$ is the period for the elliptic curve.
Comparing the curves in \eccan\ and \SWcurve, we find that
\eqn\omeq{
\om=\left({\tE_4\over f(u,m_1,m_2,\tau)}\right)^{1/4} \ .}
The coordinate $\phi$ is then found by integrating $\om$ over $u$,
with the result
\eqn\phiint{
\phi~=~ -{1\over2\pi i}\int{du}\left({\tE_4\over
f(u,m_1,m_2,\tau)}
\right)^{1/4}\ ,}
while the dual coordinate $\phi_D$ is found by integrating
$\om\ttau$ over $u$,  leaving
\eqn\phidint{
\phi_D ~=~ -{1\over2\pi i}\int{du}\left({\tE_4\over
 f(u,m_1,m_2,\tau)}\right)^{1/4} \ttau \ .}
The variable $\ttau$ is solved by equating the series expansion of
${\tE_4}^3/{\tE_6}^2$ with
${{f(u,m_1,m_2,\tau)}^3 \over {g(u,m_1,m_2,\tau)}^2}$
and then inverting the series.  The pre-potential, ${\cal F}$
is then obtained by integrating $\phi_D (\phi,\tau;m_i)$ with
respect to $\phi$.

\newsec{BPS States of E-string With Low Winding Numbers}

In this section we apply the methods of the previous section
to the problem of computing the BPS states of E-strings with
low winding numbers.

\subsec{The singly wound E-string}
Using the equations \phiint\ and \phidint\ we find that, to
 leading order in $1/u$ and up to an integration
constant,
$\phi=-{1\over 2\pi i}\log u$ and the difference of
the  dual coordinate with $\tau\phi$ is
\eqn\phidoneinst{\eqalign{
\phi_{D}-\tau\phi&=-{54{\vt_4}^4\over4\pi^2\Delta\chi(m_1,m_2)}
\Big[
{\vt_3}^6\vt_3(m_1)\vt_3(m_2)+{\vt_2}^6\vt_2(m_1)\vt_2(m_2)
\cr &\qquad\qquad+
{\vt_4}^6\vt_4(m_1)\vt_4(m_2)\Big]e^{2\pi i \phi}
 ~+~{\rm O}(e^{4\pi i\phi}) \ ,}}
where $\Delta={E_4}^3-{E_6}^2=1728\eta^{24}$ and
\eqn\chieq{
\chi(m_1,m_2)={\vt_3}^2\vt_3(m_1)\vt_3(m_2)-
{\vt_2}^2\vt_2(m_1)\vt_2(m_2)+{\vt_4}^2\vt_4(m_1)\vt_4(m_2) \ .}
More details of this calculation can be found in  appendix A.
The term inside the square brackets is the contribution of the $E_8$
lattice with two non-zero Wilson lines.  To match \phidoneinst\ to  
the
results of \refs{\KMV \OGanE}, we should multiply this result by
$16\chi(m_1,m_2)
\eta^{12}/{\vt_4}^4$, which is equivalent to changing
the integration constant in \phiint.  This shift in the integration
constant means that each term in the $phi$-expansion of
$\phi_D-\tau\phi$ will transform with a certain modular phase, the
implications of which will be discussed in the next subsection.
The Yukawa coupling is now found by
taking two $\phi$ derivatives on $\phi_D$ and can be written as
\eqn\Yuk{ {\partial^2\phi_D\over \partial\phi^2}~=~
\sum_{n=1}^\infty ~n^3~ Z_n ~ e^{2\pi i n\phi} \ . }
Therefore, the first instanton
contribution to the Yukawa coupling, which counts the holomorphic
curves of degree one on the del Pezzo surface, is
\eqn\degone{ Z_1~=~ {1\over\eta^4(\tau)}~
{ P(m_i;\tau)\over\eta^{8}(\tau)}~=~ {\chi_{1,0}\over\eta^4(\tau)} \
,}
where $P(m_i;\tau)$ is the $E_8$ lattice contribution for generic
$m_i$:
\eqn\Epart{P(m_i;\tau) ~=~ {1 \over 2}~\bigg[\sum_{\ell =1}^4~
\prod_{j=1}^8~ \vartheta_\ell (m_j)~\bigg] \ . }
The function  $\chi_{1,0}$ is the $E_8$ level one character.
We thus recover the proper E-string partition function \ZoneBnine.

\subsec{The doubly wound E-string; U(2) Yang-Mills on $\half K3$}

We could find the contribution to the doubly wound E-string by
carrying out the $u$-expansions in \phiint\ and \phidint\ to next  
order.
However, this turns out to be laborious.  Instead we will utilize
the recursion relation \recursion\  and the existence of gaps.
{}From the recursion relation we see that
\eqn\Ftwo{Z_2 ~=~ \coeff{1}{8}~Z_2^{(0)} ~+~ \coeff{1}{24}~E_2 ~
(Z_1)^2  \ , }
where $Z_2^{(0)}$ has no $E_2$ dependence. The first
term in the right hand side of \Ftwo\ is modular invariant with
weight $-2$, except for a modular phase coming from the non-zero  
$m_i$.
Recall that $\vt$-functions with non-zero arguments pick up extra
phases under modular transformations.
Hence, under a modular transformation $P(m_i;\tau)$ transforms as:
\eqn\modanom{
P\left({m_i\over c\tau+d}\ ;{a\tau+b\over c\tau+d}\right)=
(c\tau+d)^4~\exp\left({i\over c\tau+d}\sum_i m_i^2\right)~
P( m_i;\tau) \ .}
The modular ``phase'' can be compensated for in the Yukawa coupling
if, under the modular transformation, $\phi$ is shifted by
$\sum m_i^2/(c\tau+d)$.  Indeed, one should recall from the
last sub-section that the origin of the modular phase is
precisely because of the choice of a ``constant of integration''
in $\phi$.   This means that the modular phase for $Z_2^{(0)}$
has to be the  same as the modular phase for $(Z_1)^2$.  More  
generally,
this is one way of seeing that the $n^{th}$-instanton contribution,
$Z_n$, must transform as in \modprop.

We now look for generic expressions for $Z_2^{(0)}$ that have the
correct modular weight and modular phase.

Notice that $P(n m_i; n \tau)$ has the same modular phase as
$(P(m_i; \tau))^n$.  Ignoring the modular phase,
$P(n m_i; n \tau)$ is obviously not a modular form of
$SL(2,\ZZ)$, but it can be viewed as a modular form of weight four
for the subgroup $\Gamma_1(n)$ of $SL(2,\ZZ)$.  This subgroup
is comprised of matrices
$\left(\matrix{a&b\cr c&d}\right)$ where $ad-bc=1$ and $a,d= 1\ {\rm
mod}\ n$
and $c=0\ {\rm mod}\ n$.
If $n$ is prime, then under the full modular group $P(n m_i;n\tau)$
 can be transformed to $P(m_i;\tau/n+\ell/n)$ where $\ell$ is an
integer with $0\le \ell <n$.  $P(m_i;\tau/n)$ is not a $\Gamma_1(n)$
form,
but it is a form for the smaller subgroup $\Gamma(n)$, which has the
additional
requirement that $b=0\ {\rm mod}\ n$.

Hence, for the second instanton, we take the Ansatz for $Z_2^{(0)}$:
\eqn\psiIIans{\eqalign{
Z_2^{(0)}&={1\over\eta^{24}}~\Big(f(\tau)~P(2m_i;2\tau)~+~
{1\over 16\tau^6}~f(-1/\tau)~ P(m_i;\tau/2)\cr
&\qquad\qquad  ~+~{1\over 16(\tau-1)^6}~f(-1/(\tau-1))~
P(m_i;\tau/2+1/2)\Big) \ .}}
In order that $Z_2^{(0)}$ have modular weight $-2$, $f(\tau)$ must
have weight
six.  The factors of $1/16$ appear from the transformation of
$P(2m_i;2\tau)$
while the factors of $\tau^{-6}$ come from transforming $f(\tau)$.
Since the instanton expansion has integer coefficients and since
$Z_2$ is
a modular form (up to the modular phase), the function
$f(\tau)$ must be a weight six form of
$\Gamma_1(2)$.  The $\G_1(2)$ forms are generated by the weight two
form
${\vt_3}^4+{\vt_4}^4$ and the weight four form ${\vt_2}^8$.  Hence
$f(\tau)$
is determined up to two coefficients and has the form
\eqn\feq{
f(\tau)=\left({\vt_3}^4+{\vt_4}^4\right)\left(
a_1\left({\vt_3}^4+{\vt_4}^4\right)^2+a_2{\vt_2}^8\right) \ ,}
while the other terms in the orbit are given by
\eqn\forbit{\eqalign{
f(-1/\tau)&=-\tau^6\left({\vt_3}^4+{\vt_2}^4\right)\left(
a_1\left({\vt_3}^4+{\vt_2}^4\right)^2+a_2{\vt_4}^8\right)\cr
f(-1/(\tau-1))&=-(\tau-1)^6\left({\vt_4}^4-{\vt_2}^4\right)\left(
a_1\left({\vt_4}^4-{\vt_2}^4\right)^2+a_2{\vt_3}^8\right) \ .}}

The $q$ expansion of $Z_2$ has the form
$Z_2=q^{-1}+ {\rm O}(q)$, that is, there is a gap for the $q^0$ term.
The coefficients $a_1$ and $a_2$ in \psiIIans\ can be adjusted to
match
the first two terms in the expansion.  The unique choice for these
coefficients
is $a_1={1\over12},\ a_2=-{1\over12}$.  Thus, $f(\tau)$ is
\eqn\feqII{ f(\tau) ={1\over3}{\vt_3}^4{\vt_4}^4 \big(
{\vt_3}^4+{\vt_4}^4\big)  \ .}
We have also checked this result by making the direct computation
of the second order term in the pre-potential.

%
%

The  ${Z_1}^2$ term that appears in $Z_2$ can be linearized in terms
of
invariant Weyl orbits.  The number of inequivalent Weyl orbits
is equal to the number of level 2 $E_8$ characters, which
is three.  As with $K3$, the three orbits are the trivial, even and
odd, with
\eqn\levIIorb{\eqalign{
P_0(m_i,\tau)&=P(2m_i;2\tau) \ , \cr
P_{even}(m_i,\tau)&=\half\Big(P(m_i;\tau/2)+ P(m_i;\tau/2+1/2)\Big)
- P(2m_i;2\tau) \ , \cr
P_{odd}(m_i,\tau)&=\half\Big(P(m_i;\tau/2) -
P(m_i;\tau/2+1/2)\Big) \ ,}}
where the trivial orbit is the contribution from lattice vectors that
are twice another lattice vector, the even orbit is the contribution
from lattice
vectors that have length squared $0\ {\rm mod}\ 4$ and which are not
twice another lattice vector, and the odd orbit is the contribution
from lattice vectors with length squared $2\ {\rm mod}\ 4$.
The ${Z_1}^2$ term is basically the tensor product of
two level one characters, hence this term is expressible in terms
of the level two Weyl orbits.  The result has the very simple form
\eqn\Fonesq{\eqalign{
{Z_1}^2 ~=~ {1\over\eta^{24}(\tau)}\Big(P_0(0,\tau)~
P_0(m_i,\tau) ~+~ & {1\over135}P_{even}(0,\tau)~P_{even}(m_i,\tau)
\cr ~+~ & {1\over120} P_{odd}(0,\tau)~P_{odd}(m_i,\tau)
\Big)\ .}}
Notice that \Fonesq\ has the structure of a reducible connection of
$U(2)\to U(1)\times U(1)$.  The $P_\a$ denote the three types of
$U(1)$
fluxes on the del Pezzo surface, with a $P_\a$ factor for each
$U(1)$.

The $Z_2^{(0)}$ term in \psiIIans\ can also be expressed in terms of
the invariant
Weyl orbits.  Hence, the entire $Z_2$ term can be written as
\eqn\psiIIweyl{
Z_2=Z_0(\tau)P_0(m_i,\tau) + Z_{even} P_{even}(m_i,\tau)+
Z_{odd }P_{odd}(m_i, \tau) \ ,}
where
\eqn\Zres{\eqalign{
Z_0(\tau)&={1\over24\eta^{24}}\left(\hE_2P_0(0,\tau)+\left(
{\vt_3}^4{\vt_4}^4-{1\over8}{\vt_2}^8\right)
 \left({\vt_3}^4+{\vt_4}^4\right)\right)\cr 
Z_{even}(\tau)&={1\over24\eta^{24}}
\left({1\over135}\hE_2P_{even}(0,\tau)-{1\over8}{\vt_2}^8
\left({\vt_3}^4 + {\vt_4}^4\right)\right)\cr 
Z_{odd}(\tau)&={1\over24\eta^{24}(\tau)}
\left({1\over120}\hE_2P_{odd}(0,\tau)- {1\over8}{\vt_2}^4
E_4\right).}}
Thus, we obtain the partition function
of $N=4$ $SU(2)$ Yang-Mills on $\half K3$, corresponding
to three inequivalent types of `t Hooft flux on the $E_8$
part of  $H_2(\half K3)$ and subject to
the K\"ahler limit described in section 3.
We now learn from \Zres\ that  the
holomorphic anomaly in this case is
\eqn\holanomag{
\overline \partial Z_\lambda ={i\over 2 \pi \tau_2^2}
{C_\lambda\over\eta^{24}(\tau)} P_\lambda (0,\tau),}
where
\eqn\Clameq{
C_0=1\qquad\qquad C_{even}={1\over135}\qquad\qquad  
C_{odd}={1\over120}.}
$P_\lambda(0,\tau)$ corresponds to the class of
$U(1)\subset SU(2)$ fluxes in  $H_2(\half K3)$ corresponding to the
't Hooft flux denoted by $\lambda$.  Note that there are $2^8$  
distinct
`t Hooft fluxes and that ${C_\lambda}^{-1}$ is the number of fluxes  
for
each class $\lambda$.  We thus conclude that the weight factors  
$C_\lambda$
are the
inverses of the number of `t Hooft fluxes in each class.\foot{The  
holomorphic
anomaly for $SU(2)/Z_2$ on ${\bf P}^2$ has a similar dependence on  
the
flux degeneracy.  In this case there are two classes, each with a  
prefactor
of $1$.  These correspond to the two possible choices of 't Hooft  
flux,
one trivial and one nontrivial.}
Each $Z_\lambda$ is the contribution to the $SU(2)/Z_2$ partition  
function
for one particular flux in the class $\lambda$.  To find the
full partition function, one should sum over each flux in the class.
This will cancel out the degeneracy factors $C_\lambda$ in  
\holanomag,
and so  for the full
partition function each class
comes with an equal weight in the anomaly equation and is a natural
setup to explain the above result.

The three functions $P_0(m_i,\tau), P_{even}(m_i,\tau)$ and
$P_{even}(m_i,\tau)$ can also be written in terms of level
two characters of $E_8$. To determined the characters we note that
\eqn\ebranch{
 E_8^{(1)} \times E_8^{(1)} \ = \ ({\it  Ising \ model})~
E_8^{(2)} \ ,}
where the superscript in parentheses denotes the level of the
current algebra.
The branching functions are given by the chracters of the Ising
models.  After some straight forward algebra it is easy to check
that the level two characters for $E_8$ are given by
\eqn\newchar{\eqalign{
 \chi_{248} \ = \  &{1\over 2}{\eta(\tau) \over\eta(2 \tau)}
\Biggl( {(P(m_i,\tau))^2\over\eta(\tau)^{16 }}-{P(2m_i,2\tau)\over
\eta(2\tau)^8 } \Biggr )  \cr
 \chi_{0} \ = \ &{1\over 2}{\eta(\tau) \over\eta( \tau/2)}\Biggl(
{P(m_i,\tau)^2\over\eta(\tau)^{16
}}-{P(m_i/2,\tau/2)\over\eta(\tau/2)^8 } \Biggr )  +
\cr  & {1\over 2}{\eta(\tau+1) \over\eta( \tau/2+1/2)}\Biggl(
{P(m_i,\tau)^2\over\eta(\tau+1)^{16
}}-{P(m_i/2+1/2,\tau/2+1/2)\over\eta(\tau/2+1/2)^8 } \Biggr )
\cr  \chi_{3875} \ = \ &{1\over 2}{\eta(\tau) \over\eta( \tau/2)}
\Biggl( {P(m_i,\tau)^2\over\eta(\tau)^{16 }}-{P(m_i/2,\tau/2)\over
\eta(\tau/2)^8 } \Biggr )  -
\cr & {1\over 2}{\eta(\tau+1) \over\eta( \tau/2+1/2)}\Biggl(
{P(m_i\tau)^2\over\eta(\tau+1)^{16 }}-{P(m_i/2+1/2,\tau/2+1/2)\over
\eta(\tau/2+1/2)^8 } \Biggr ) \ . }}
Inverting these equations we have
\eqn\inver{\eqalign{{P_0(m_i,\tau )\over \eta(\tau)^8} \ = \
&{\eta(2\tau)^8 \over \eta(\tau)^8}\ (\chi_0 \chi_{1,1}-
\chi_{248} \chi_{1,2}+\chi_{3875} \chi_{2,1})
\cr
{P_{even}(m_i,\tau )\over \eta(\tau)^8} \ = \  &
{\eta(\tau/2)^8 \over \eta(\tau)^8} ((\chi_0  +\chi_{3875} )
(\chi_{1,1}+\chi_{2,1})+\chi_{248} \chi_{1,2})-
\cr & {\eta(\tau/2+1/2)^8 \over \eta(\tau)^8}
((\chi_0  -\chi_{3875} )(\chi_{1,1}-\chi_{2,1} )+\chi_{248}
\chi_{1,2})+
\cr &{\eta(2\tau)^8 \over \eta(\tau)^8}\ (\chi_0 \chi_{1,1}-
\chi_{248} \chi_{1,2}+\chi_{3875} \chi_{2,1})
\cr {P_{odd}(m_i,\tau )\over \eta(\tau)^8} \ = \ &
{\eta(\tau/2)^8 \over \eta(\tau)^8}
((\chi_0  +\chi_{3875} )(\chi_{1,1}+\chi_{2,1})+\chi_{248}
\chi_{1,2})+
\cr & {\eta(\tau/2+1/2)^8 \over \eta(\tau)^8}
((\chi_0  -\chi_{3875} )(\chi_{1,1}-\chi_{2,1})+\chi_{248}
\chi_{1,2})\ ,}}
where  $ \chi_{1,1}, \chi_{1,2}$ and $\chi_{2,1} $ are the characters
of the Ising model.

One can now substitute this into \Zres\ and \psiIIweyl, and expand
to obtain an explicit expression for $Z_2$ in terms of level
$2$ characters of $E_8$ and ``branching functions'' of some
conformal field theory with central charge $c = 9/2$.
Similarly, for the $Z_n$ case we end up with an $E_8$ current
algebra at level $n$, 4 transverse bosonic oscillator and a leftover
conformal system with central charge $c=12n-4-{248n\over n+30}$.

Note that the combinations of $\eta$-functions that multiplies
the $E_8$ characters can be written in terms $SU(2)_1$ characters
$\chi^{su(2)}_0$ and $ \chi^{su(2)}_1$:
\eqn\repsu{ {\eta(2\tau)^8 \over \eta(\tau)^8} \ = \ (\chi^{su(2)}_0-
\chi^{su(2)}_1)^4 \ .}

\subsec{Expressions for $Z_3$ and $Z_4$}

We now continue the foregoing computations to obtain the partition
functions of the three- and four-times wrapped E-string,
corresponding to $SU(3)$ and $SU(4)$ Yang-Mills on $\half K3$.
In this section we give the explicit formulae for  $Z_3$ and $Z_4$,
and discuss the Ansatz that is appropriate at higher orders.
Even though the techniques we have discussed yield answers for all
$n$
it becomes more and more laborious to write the explicit form for
higher $n$'s.
As we have seen, the higher functions $Z_n$
have the form $Z_n(m_i; \tau) = g_n(m_i; \tau)/\eta^{12n}$, where
 $g_n(m_i; \tau)$ is a modular form of weight $6n-2$
(but with a modular phase of the form \modprop).
The functions  $g_n(m_i; \tau)$ have two types of component:
$E_8$ root lattice terms, containing the $m_i$ dependence and
coefficient functions for each of these terms.
To get $Z_n$ for $n \ge 3$ we once again make an Ansatz and use
the recurrence relation \recursion\ and the gap condition
\gapcond.  Here we will describe the first Ansatz that we made,
present the results, and in the next sub-section we will
justify why the Ansatz is correct, and improve  upon it.

The $E_2$-dependent parts of $g_n$ are determined by \recursion,
and so we need only make an Ansatz for the $E_2$-independent part.
As in the last section,  we start with the $E_8$ building blocks:
$P(n m_i;n \tau)$,
and $P(m_i;(\tau + j)/n)$, $j=0,1,\dots,(n-1)$, since these
functions have the proper periodicity, and the correct
$m_i$-dependent phases under modular inversion.  The coefficient
functions must be modular forms of $\Gamma(n)$, and have weight
$6(n-1)$.  Moreover, since $P(n m_i;n \tau)$ is invariant under
$\tau \to \tau +1$, its coefficient must be a modular form of
$\Gamma_1(n)$.  We therefore make an Ansatz for this coefficient
function, and then determine all the other terms by taking the
orbit under the full modular group.  For $g_2$, such an Ansatz
was sufficient, but not for the higher $g_n$.  One must also allow
$E_8$-building blocks of the form:
\eqn\genbuildbk{ \prod_k ~ P(n_k m_i; n_k \tau) ~~
\prod_\ell ~ P(m_i;  (\tau + j_\ell)/n_\ell) \ ,}
where $\sum n_k + \sum n_\ell = n$. When all such terms are included
the system is highly overdetermined for almost all the coefficients,
but there are apparently one or two underdetermined constants in the
final result.  Upon closer inspection,
it turns out that these undetermined coefficients actually
multiply an expression that is identically zero by virtue
of some peculiar identity.

We find the following results for $Z_3$:
\eqn\fthreeres{\eqalign{Z_3 ~=~  {1 \over 864~\eta^{36}}~
\Big[& ~A_0(\tau)~ P(3 m_i; 3 \tau) ~+~ A_1(\tau)~ P\big(m_i;
{\tau \over 3}\big)  ~+~ A_1(\tau+1)~  P\big(m_i;
{(\tau + 1)\over 3}\big)  \cr & ~+~ A_1(\tau+2)~  P\big(m_i;
{(\tau + 2)\over 3}\big)  ~-~ 3~E_4(\tau)~ (P(m_i;\tau) )^3
\Big] \cr &  ~+~ {1 \over 6} ~E_2(\tau) ~Z_1(m_i;\tau)
Z_2(m_i;\tau) ~-~  {1 \over 288} ~(E_2(\tau))^2 ~
(Z_1(m_i;\tau))^3 \ ,}}
where
\eqn\Afunctions{\eqalign{A_0(\tau)  ~=~  & 20~{\eta^{36}(\tau)
\over  \eta^{12}(3 \tau) } ~+~ 972~\eta^{24}(\tau) \ , \cr
A_1(\tau)  ~=~ & 12~\bigg[~15~{\eta^{36}(\tau) \over
\eta^{12}(\tau/3) } ~+~ \eta^{24}(\tau)~\bigg] \ .}}
For $Z_4$ we obtain:
\eqn\ffourres{\eqalign{Z_4 ~=~  {1 \over 1152~\eta^{48}}~
\Big[& ~B_0(\tau)~ P(4 m_i; 4 \tau) ~+~ B_1(\tau)~ E\big(m_i;
{\tau \over 4}\big)  ~+~ B_1(\tau+1)~  E\big(m_i;
{(\tau + 1)\over 4}\big)  \cr & ~+~ B_1(\tau+2)~  E\big(m_i;
{(\tau + 2)\over 4}\big)  ~+~ B_1(\tau+3)~  E\big(m_i;
{(\tau + 3)\over 4}\big) \cr & ~+~ B_2(\tau)~  E\big(2 m_i;
\tau + \half \big) ~+~  E_6(\tau)~ \eta^{24}(\tau)~ P(2 m_i;
\tau ) ~\Big] \cr &  ~+~ {1 \over 12}~E_2(\tau) ~(3~ Z_1(m_i;\tau)~
Z_3(m_i;\tau) ~+~ 2~(Z_2 (m_i;\tau))^2) \cr &~-~
{1 \over 36} ~((E_2(\tau))^2 ~+~
E_4(\tau)) ~ (Z_1(m_i;\tau))^2 ~Z_2(m_i;\tau) \cr & ~+~
{1 \over 2592} ~((E_2(\tau))^3 ~+~ 3~E_2(\tau)~E_4(\tau) ~-~
E_6(\tau)) ~(Z_1(m_i;\tau))^4\ ,}}
where
\eqn\Bfunctions{\eqalign{B_0(\tau)  ~=~  & \vartheta_3(0|2 \tau)^4~
\vartheta_4(0|2 \tau)^{24}~ \big[32~\vartheta_3(0|2 \tau)^8  \cr &
\qquad \qquad ~-~
20~\vartheta_3(0|2 \tau)^4~\vartheta_4(0|2 \tau)^4  ~-~
\vartheta_4(0|2 \tau)^8~\big] \ , \cr
B_1(\tau)  ~=~ & - {1 \over 2^{26}}~\vartheta_3(0|\tau/2)^4~
\vartheta_2(0|\tau/2)^{24}~ \big[11~\vartheta_3(0|\tau/2)^8 \cr &
\qquad \qquad  ~+~
22~\vartheta_3(0|2 \tau)^4~\vartheta_4(0|\tau/2)^4  ~-~
\vartheta_4(0|\tau/2)^8~\big] \ , \cr
B_2(\tau)  ~=~ &  {1 \over 16}~ \vartheta_2(0|2 \tau)^4~
\vartheta_4(0|2 \tau)^{24}~ \big[11~\vartheta_2(0|2 \tau)^8  \cr &
\qquad \qquad ~+~
22~\vartheta_2(0|2 \tau)^4~\vartheta_3(0|2 \tau)^4  ~-~
\vartheta_3(0|2 \tau)^8~\big]  \ .}}

\subsec{The Structure of the Partition Functions: Weyl Orbits}

To understand and generalize the Ans\"atze above, we note several
properties of the partition functions.
First, as functions of the $m_i$, the parammeters of the
torus are doubly periodic under translations  $m_i \to m_i +
\alpha_i$ and $m_i \to m_i + \tau \alpha_i$, where $\alpha_i$
is a root of $E_8$.  It follows that the $m_i$ dependence of
the partition functions must appear as a sum over a
scaled version of the $E_8$-root lattice.  The precise scale
is set by the modular phase and the modular invariance.
We also know that the function $Z_n(m_i;\tau)$ must contain a part
that is $Z_1(n m_i;n \tau) = P(n m_i; n \tau)$, coming from
the multiple windings of the base of the elliptic fibration of
$\half K3$.  This leads fairly unambiguously to a generic
$E_8$ root lattice term in $g_n$ of the form:
\eqn\genchar{P_{n, \lambda}(m_i; \tau) ~=~  \sum_{w \in W(E_8)}
\sum_{\alpha \in
\Lambda(E_8)}~ q^{{1 \over2 n} ~(\lambda + n \alpha)^2}~
e^{2 \pi i  \vec m \cdot w (\vec \lambda + n \vec \alpha)} \ ,}
where $\Lambda(E_8)$ and $W(E_8)$ are, respectively, the root
lattice and Weyl group of $E_8$.  The sum over the Weyl group
is included since the rational curves of
$\half K3$ have an $E_8$ Weyl invariance, and so the
partition functions must have this invariance.
While the individual $P_{n, \lambda}$ are not modular
invariant, they transform into one another with a modular
weight of $4$ and a modular phase of the form \modprop.

To count the number of independent functions $P_{n, \lambda}$,
one views them as the partition functions
of the diagonal $U(1)^8$ in a product of $n$ copies $E_8$, or
equivalently, one thinks of them as the $U(1)^8$ part
of the partition function of the a representation of the
current algebra  $E_8^{(n)}$.  The number of independent functions,
$P_{n, \lambda}$, is thus equal to the number of characters
of $E_8^{(n)}$.  For $n=1,2,3,4$ this number is $1,3,5,10$.

The function $g_n$ is then given by:
\eqn\weightsum{g_n(m_i;\tau) ~=~ \sum_{\lambda}~Z_\lambda(\tau)~
P_{n, \lambda}(m_i;\tau) \ ,}
where the functions $Z_\lambda(\tau)$ are modular forms
of weight $6(n-1)$ in $\Gamma(n)$.  In particular $Z_0$ must
be a modular form of $\Gamma_1(n)$.

One can now understand the role of the general building blocks
of the form \genbuildbk.   There are $n+1$ functions
of the form $P(n m_i;n \tau)$, and $P(m_i;(\tau + j)/n)$,
 but for $n \ge 3$ there are more than $n+1$ functions
$P_{n, \lambda}$.  We therefore need to find ways to weight
the characters of independent Weyl orbits in different ways.
This can be done explicitly as in \weightsum, or implicitly using
expressions like \genbuildbk.

Rewriting the partition functions as weighted sums over
Weyl orbits is very natural from the Yang-Mills perspective, and
is also natural in finding universal properties of the
characters \MNWa.  One can thus rewrite the partition function
$Z_3$ as follows.

Introduce the functions:
\eqn\suthree{\eqalign{h_0(\tau) ~=~ & \sum_{n_1,n_2 =
- \infty}^\infty~ q^{n_1^2 + n_2^2 - n_1 n_2} \ , \qquad
h_1(\tau) ~=~ \half(h_0(\tau/3) - h_0(\tau)) \ , \cr
h_2(\tau) ~=~ & {\eta^9(\tau) \over \eta^3(3 \tau)} \ ; \qquad
h_3(\tau) ~=~ 27~{\eta^9(3 \tau) \over \eta^3(\tau)} \ .}}
Note that: (i) $h_0$, $h_2$ and $h_3$ are modular forms of
$\Gamma_1(3)$ of weights $1,3$ and $3$ respecively; (ii)
$h_0$ and $h_0(\tau/3)$ are, respectively, the partition
functions of the root and weight lattice of $SU(3)$; and
(iii) one has $h_0^3 = h_2 + h_3$.  Define the functions
$Q_j$ as follows:
\eqn\threedecomp{\eqalign{Q_0(m_i,\tau) ~=~ & P(3 m_i;3 \tau) ~=~
P_{3,\lambda = 0} \ , \cr Q_j (m_i,\tau) ~=~ & {1 \over 3}~\Big(
P \big(m_i;{\tau \over 3} \big) ~+~  \omega^{j-1}~P \big(m_i;
{(\tau+1) \over 3}  \big) ~+~ \omega^{2(j-1)} P \big(m_i;
{(\tau + 2) \over 3} \big) \Big)\ , \cr \
& \hskip 8cm j = 1,2,3 \ , \cr
Q_4 (m_i,\tau) ~=~ &  \sum_{\beta \in \Delta(E_8)}
\sum_{\alpha \in \Lambda(E_8)}~ q^{{1 \over 6} ~(\beta + 3
\alpha)^2}~
e^{2 \pi i  \vec m \cdot (\vec \beta + 3 \vec \alpha)}
\ ,}}
where $\omega = e^{2 \pi i/3}$, and $\Delta(E_8)$ and  $\Lambda(E_8)$
are, respectively, the roots and root lattice of $E_8$.
Note that $Q_j$ represents the projection of the root lattice onto
those
vectors of $(length)^2 \equiv 2(j-1) \ mod \ 6$.  The functions
$P_{3,\lambda}$ are then: $Q_0, Q_4, Q_3,(Q_1 - Q_0)$ and
$(Q_2 - Q_4)$.

One then has:
\eqn\fthreedecomp{\eqalign{Z_3(m_i; \tau) ~=~  {1 \over
864~\eta^{36}}~&\Big[~h_2^2~\big[(17~ h_2^2 + 6 h_2 h_3 -
27 h_3^2) ~+~ 12~E_2~ h_0 h_2 ~+~ 3~E_2^2~ h_0^2~\big]~Q_0
\cr & ~+~ h_3~\big[(\coeff{2}{3}~ h_2^3 +
57~h_2^2 h_3 +204 h_2 h_3^2 + 153 h_3^3) \cr  &  ~-~
12~E_2~h_0~(\coeff{1}{9}~h_2^2 + \coeff{7}{3}~
h_2 h_3 + 3~h_3^2)  ~+~ 3~E_2^2~h_0^2~ (\coeff{2}{9} h_2 + h_3)~
\big]~Q_1 \cr & ~+~ h_3 h_1^2~\big[~h_0~(24~ h_2^3 +
153~h_2 h_3 + 153~h_3^2) \cr  &  ~-~ 4~E_2~h_0^2~(4~h_2 +
9~h_3)  ~+~  E_2^2~ (\coeff{8}{3}~h_2 + 3~h_3)~\big]~Q_2
\cr & ~+~ h_3 h_1~\big[~h_0^2~(5~ h_2^3 +
102~h_2 h_3 + 153~h_3^2) \cr  &  ~-~ \coeff{4}{3}~E_2~(5~h_2^2 +
30 h_2 h_3 + 27~h_3^2)  ~+~ \coeff{1}{3}~E_2^2~h_0~(5~h_2 +
9~h_3)~\big]~Q_3
\cr & ~-~ h_2^2  h_1^2 ~\big[~h_0~(h_2  + 9~h_3 ) ~-~
E_2^2~\big]~Q_4 ~\Big] \ .}}
In terms of the Weyl orbits, the function $Z_3$ decomposes as:
\eqn\fthreeother{\eqalign{Z_3(m_i; \tau) ~=~ {1 \over
864~\eta^{36}}~ &\Big[~\big[(17~ h_2^4 + \coeff{20}{3} h_2^3 h_3 +
30 h_2^2 h_3^2 + 204 h_2 h_3^3 + 153 h_3^4) \cr &  ~+~
12~E_2~ h_0 (h_2^3 - \coeff{1}{9} h_2^2 h_3  - \coeff{7}{3}
h_2 h_3^2 - 3 h_3^3) \cr & ~+~ 3~E_2^2~ h_0^2(h_2^2  +
\coeff{2}{9} h_2  h_3 + h_3)^2~\big]~P_{3,\lambda_0}
\cr & ~-~ h_1^2~\big[~h_0~(h_2^3 - 15~ h_2^3 h_3 -
153~h_2 h_3^2 + 153~h_3^3)   \cr & ~-~ 4~E_2~h_0^2~(4~h_2 h_3
+ 9~h_3^2)   \cr & ~-~ E_2^2~ (h_2^2 + \coeff{8}{3}~h_2 h_3 +
3 h_3^2)~\big]~P_{3,\lambda_1}  \cr &  ~+~
h_3 h_1~\big[~h_0^2~(5~ h_2^3 +
102~h_2 h_3 + 153~h_3^2)   \cr  &  ~-~
\coeff{4}{3}~E_2~(5~h_2^2 + 30 h_2 h_3
+ 27~h_3^2)  \cr  &  ~+~
\coeff{1}{3}~E_2^2~h_0~(5~h_2 +
9~h_3)~\big]~P_{3,\lambda_2}  \cr & ~+~
h_3~\big[(\coeff{2}{3}~ h_2^3 +
57~h_2^2 h_3 +204 h_2 h_3^2 + 153 h_3^3) \cr  &  ~-~
12~E_2~h_0~(\coeff{1}{9}~h_2^2 + \coeff{7}{3}~
h_2 h_3 + 3~h_3^2)   \cr & ~+~
3~E_2^2~h_0^2~ (\coeff{2}{9} h_2 + h_3)~
\big]~P_{3,\lambda_3} \cr & ~+~
h_3 h_1^2~\big[~h_0~(24~ h_2^3 +
153~h_2 h_3 + 153~h_3^2) \cr  &  ~-~ 4~E_2~h_0^2~(4~h_2 +
9~h_3)  ~+~  E_2^2~ (\coeff{8}{3}~h_2 + 3~h_3)~\big]~
P_{3,\lambda_4}~\Big] \ .}}

\newsec{The String Interpretation}

We have seen how to compute the number of BPS states for
E-strings wrapped $n$-times around a circle.  In this section
we would like to examine the nature of the low-energy modes
propagating on the E-string.  Before doing this, it would be helpful
to review the relationship between multi-wound and singly-wound
heterotic strings.  It is natural to compare the E-string to the  
heterotic
string since the heterotic string arises by wrapping an M5 brane
around $K3$, whereas E-string arises by wrapping an M5 brane around
$\half K3$.

\subsec{Multi-wound strings in general}

It is one of the remarkable properties of M5 branes that when we
wrap $n$ of them around $K3 \times T^2$ we simply obtain the
$n$-wound heterotic string on the torus:  One has started out with
$n$ independent objects, but the process of wrapping them results
in the Hilbert space of a single string.  The multi-wrapping has
effectively resolved itself into a concatenation of strings.
As is well known,  the structure of the Hilbert space for
multi-wound heterotic strings can be easily obtained from the
Hilbert space of singly wound strings.  In particular if we wish to  
find
BPS states for the heterotic string with momentum $p$ and
winding number $n$ we should consider
$$pn=N_L-1$$
For a singly-wound string we have $p=N_L-1$, while for a string
wrapped $n$ times we have
$$p={N_L-1\over n}.$$
Given that $p$ is an integer, we obtain  states
with  oscillator numbers that are  a multiple $n$ of the basic unit.

Thus, thinking of the heterotic string as arising from an
M5-brane wrapping $K3$, we see that a ``doubled'' heterotic string
coming from two M5-branes wrapping $K3$ is in some sense not a new
object but is the same as the original heterotic string merely
doubly wound.  Moreover, the BPS states of the doubly wound
heterotic string are obtained from those of the singly-wound
string by scaling the torus and doing appropriate projections.
We interpret this as being
due to the mild interaction between contiguous heterotic strings.
However, for other types of strings we may not find such simple  
behavior
if there are strong dynamics between strings (and their parent
M5-branes) as they get close to one another.

It is also true that if a heterotic string has some level one group  
$G$
current algebra as a symmetry, then
the spectrum of BPS states with winding number $n$ has a natural
action under the level $n$  current algebra.  This is so because
the level one $G$ currents
get projected onto the subset of modes that are $0$ mod $n$
in the $n$-winding sector.
It is elementary to verify that such modes, $J^i_{p n}$ ,
$p \in \ZZ$, satisfy a level $n$ current algebra.

Now what do our results tell us about the E-string and its
dynamics when two or more strings lie close to one another?
For a singly wound E-string
the spectrum of BPS states in \degone\ is consistent with
 a level one $E_8$ current algebra.  Moreover, the
low energy dynamics seems to be that of a free theory, corresponding
to the position of the string in transverse space.
The string picture also readily predicts that all BPS partition  
functions
should have weight $(-2,0)$.  This is because the $n$
bound E-strings have 4 non-compact transverse modes for their
center of mass motion.  We also saw that the number of inequivalent
flux classes for $n$ five-branes is equal to the number
of inequivalent orbits of $\Gamma^8/n \Gamma^8$ under the $E_8$
Weyl group (we mod out by the $E_8$ Weyl group
because fluxes which map to each other under the  Weyl
group are diffeomorphically equivalent).  This matches the number of
level $n$ $E_8$ characters.
This suggests that the corresponding $n$ string bound state carries
a level $n$ $E_8$ current algebra.
This too agrees with the string picture
in the sense that a level $n$ current algebra appears in the
$n$-wound  heterotic string.

However, this is where the similarities end.  Namely the partition
function we have found for the BPS states
of  $n$-wound E-strings coming from $n$ M5 branes wrapping $\half K3$
cannot be viewed as coming from a simple
multi-winding of a one E-string, suggesting that there are  
non-trivial strong
dynamics between E-strings.   In particular our results
suggest that $n$ copies of an E-string can form {\it new bound  
states},
that are new strings in their own right.  We call these
$E^{(n)}$-strings.  These should be viewed as bound states of
$E^{(i)}$-strings at threshold with $i<n$.

We know that the low energy modes on the $E^{(n)}$-string seem to  
have a
level $n$ $E_8$ current algebra, along with some other left-over
degrees of freedom.  Of course, this is only for the left-movers
which couple to the right-moving ground state.  The rest of the modes
remain unknown.

\subsec{Holomorphic Anomaly and $E^{(n)}$-String}

We have already given an interpretation of holomorphic
anomaly from the viewpoint of $N=4$ Yang-Mills in terms of
reducible connections $U(n)\rightarrow U(k)\times U(n-k)$.
We also have given an interpretation of the anomaly
in terms of holomorphic curve counting.
We would also like to interpret
the anomaly from the viewpoint of the $E^{(n)}$-string.
Though a bit less precise than the other versions, this
nevertheless teaches us something about the spectrum
of the Hilbert space of $E^{(n)}$.

It should be clear from the previous discussions
that the anomaly is related to the decomposition
$E^{(n)}\rightarrow E^{(k)}+E^{(n-k)}$.  In fact we can see this as
follows: The formal argument that
$Tr\big[(-1)^F F_R^2q^L_0 {\overline q}^{\overline L_0}\big]$ is
holomorphic in $\tau$ (for non-compact bosons) relies on the
cancellation of supersymmetric states in pairs to remove all
non-holomorphic contributions.  However given the existence of a  
channel
$E^{(n)}\rightarrow E^{(k)}+E^{(n-k)}$, one expects  the Hilbert
space of the $E^{(n)}$ string to be
gapless.  When there is no gap the formal arguments for
holomorphicity can have potential anomalies, basically
because the density of states in a non-compact space may
differ between  bosons and their fermionic partners. Examples
of this were found in \CFIV.
Thus we find a natural string interpretation for the anomaly.

\newsec{Suggestions for Further Study}

We have studied the
BPS states of E-strings from various perspectives and thus have  
connected
the counting of holomorphic curves with the
partition functions of $N=4$ topological Yang-Mills on
${\half K3}$.  We have found that the
holomorphic anomaly first observed for topological strings
\BCOV\ is related to the holomorphic anomaly for  BPS
state counting  \refs{\MNWb,\MNWc} which in turn is related
to the holomorphic anomaly for topological $N=4$ Yang-Mills
on manifolds with $b_2^+=1$ \CVEW .

The BPS state partition functions can be completely determined
using four inputs: modular invariance, the holomorphic anomaly,
the gap,  and the winding number 1 partition
function $Z_1$.  It is amusing that each of these facts can perhaps
be understood from different perspectives:
The holomorphic anomaly is natural from the curve counting viewpoint.
The presence of the gap and $SL(2,Z)$ invariance are natural
from both the curve counting and the $N=4$ Yang-Mills viewpoint and  
the
partition function
$Z_1$ with a level one $E_8$ current algebra is
natural from the free string description of the single E-string.

There are many possible extensions of this work.  We have
found the partition functions for left-moving excitations
responsible for the BPS states of the E-strings.  The partition  
functions
contain a level $n$ $E_8$ current algebra, along with a computable
but not so easily
identifiable piece.  It would be nice to associate this
extra piece with a specific two-dimensional quantum field theory.


We have also computed the partition function for $N=4$ topological
$U(n)$ Yang-Mills
on $K3$, extending the results in \CVEW . The derivation
has a simple interpretation in terms of M5 branes,
but it is not rigorous, since we assumed that coincident
points of M5 branes in  target space do not contribute anything
extra to the partition function.  It would be nice to make this
statement rigorous by finding a topological deformation that
separates the M5 branes while preserving at least $N=1$
supersymmetry.  It is also natural to ask if we can extend these
results to other groups.  We probably have enough
ingredients to actually do the calculation.  It should also be
possible to extend the results for $K3$ to manifolds with
$h^{2,0} > 1$. We gave a heuristic argument in terms of M5 branes
that this should be possible.

We have computed the partition function of $N=4$ Yang-Mills
on a specific manifold with $b_2^+=1$.  There are other manifolds
(for example $P^2$ blown up at $n\not=9$ points) for which
we do not know the answer. In fact the case with $n<9$ is interesting
for counting BPS states for five-dimensional versions of E-strings
\refs{\DKV,\DMNS,\MNWa}.  The count in \KMV\ applies only to electric
M2 branes wrapped inside these del Pezzo surfaces,
whereas the N=4 Yang-Mills theory on these spaces computes the
magnetic states, which are obtained from the wrapped 5-brane.
Only for $\half K3$, which is {\it elliptic}, are these two  the  
same,
because the $SL(2,Z)$ on the fiber generates electric/magnetic  
self-duality.
We can also turn this around. In particular the computation of the  
$SU(2)$
partition function on ${\bf P}^2$ discussed in \CVEW\  (see also  
\VWmath)
gives the number of BPS bound states of two M5-branes wrapped around
$P^2\times S^1$.  Note that this partition
function is not related in a simple
way  to the partition function of a single M5 brane
wrapped around ${\bf P}^2\times S^1$.  This shows that also here
new bound strings are appearing at threshold, very much
like the E-string story.  Also the fact that the moduli
space of instantons on ${\bf P}^2$ blown up at up to 8 points
varies dramatically as we change the K\"ahler metric, implies that
the $N=4$ topological Yang-Mills partition function
on these spaces, which computes
the number of BPS states of the corresponding string,
will exhibit similar phenomena.  This is then interpreted
from the $N=2$ field theory perspective in four dimension
as the decay of BPS states as we vary the vector multiplet
moduli of $N=2$ theories.  It would thus be very interesting to
develop techniques to compute the $N=4$ topological $U(n)$
Yang-Mills on these spaces as well.

\goodbreak
\vskip1.cm
\noindent
{\it Note added:\ \ }
After completing this work K. Yoshioka has managed to prove
the formula derived in \Zres\ for the partition function of $SU(2)$  
on $\half K3$
\KY\ using \VWmath\ and results relating 
differences of Donaldson invariants to Jacobi forms \GZ.

\goodbreak
\vskip2.cm\centerline{\bf Acknowledgements}
\noindent

We would like to thank S. Katz, T. Pantev and K. Yoshioka for  
valuable
discussions.  J.M.  would like to thank the ITP at Santa Barbara for 
hospitality during the completion of this paper.
The work of J.M.,D.N. and N.W. was supported in part
by funds provided by the DOE under grant number DE-FG03-84ER-40168.
The research of C.V. was supported in part by NSF grant PHY-92-18167.

\goodbreak

\appendix{A}{Computing the BPS States For Singly Wound E-string}

Taking the curve in
\SWIIm\ and shifting $x$ by
$x\to x-{1\over3}\c u^2+{2\over3}{s_+}^2{s_-}^2u$ and rescaling $u$
to
$u\to u{\vartheta_3}^{12}$, one is left with a curve in the form of
\SWcurve\ with
\eqn\fgeq{\eqalign{
f(u,m_1,m_2,\tau)&={1\over3}\left(E_4u^4+(6A-4\c
B)u^3/\vartheta_3^{4}+
4B^2u^2/\vartheta_3^{16}\right)={1\over3}\barE_4\cr
g(u,m_1,m_2,\tau)&={2\over27}\Big(E_6u^6-27u^5+9k^2Bu^5+9\c Au^5-
6\c^2Bu^5+12\c B^2/\vartheta_3^{12}\cr
&\qquad\qquad-18ABu^4/\vartheta_3^{12}+
8B^3u^3/\vartheta_3^{24}\Big)={2\over27}\bar E_6\ .}}
The Eisenstein functions, $E_n$, can be written in terms of $k$:
\eqn\Eask{
E_4 ~=~ (1-k^2+k^4)~\vartheta_3^{8}\ ,\qquad\qquad
E_6 ~=~ \half(1+k^2)(1-2k^2)(2-k^2)~\vartheta_3^{12}\ .}
The coefficients $A$ and $B$ are given by:
\eqn\ABeq{
A={s_+}^2+{s_-}^2\qquad\qquad B={s_+}^2{s_-}^2 \ .}

To leading order in $1/u$,
$\left({\tE_4\over f(u,m_1,m_2,\tau)}\right)^{1/4}=1$, however
to lowest order $\ttau-\tau=0$. To find the next to leading order
contribution to this difference, we equate ${\tE_4}^3/{\tE_6}^2$ with
$f^3/g^2$.  Up to the next to leading order we find that:
\eqn\Eteq{
{{\tE_4}^3\over{\tE_6}^2}~=~{{E_4}^3\over{E_6}^2}+\left((12A-4\c)
{{E_4}^2\over{E_6}^2{\vt_3}^4}+(54-(18k^2+6\c^2)B-18\c A)
{{E_4}^3\over{E_6}^3}\right){1\over u}~+~{\rm O}(u^{-2})}
Taking a $u$ derivative on both sides of the equation in \Eteq\
we find
\eqn\uderiv{
{\Delta{E_4}^2\over {E_6}^3} {1\over q}{\partial \tq\over\partial u}=
-\left((12A-4\c)
{{E_4}^2\over{E_6}^2{\vt_3}^4}+(54-(18k^2+6\c^2)B-18\c A)
{{E_4}^3\over{E_6}^3}\right){1\over u^2} ~+~{\rm O}(u^{-3}),}
where $\tq=e^{2\pi i\ttau}$.  Thus,
\eqn\taudiffeq{
\ttau-\tau={1\over2\pi i\Delta}\left((12A-4\c)
{E_6\over{\vt_3}^4}+(54-(18k^2+6\c^2)B-18\c A)
{{E_4}}\right){1\over u}+ ~+~{\rm O}(u^{-2}).}
If we now substitute \taudiffeq\ and \ABeq\ into \phidint,  use
the identities
\eqn\ABrel{\eqalign{
A&~=~ 2~{{\vt_3}^2\over{\vt_2}^2}~{{\vt_2}^2\vt_3(m_1)\vt_3(m_2)-
{\vt_3}^2\vt_2(m_1)\vt_2(m_2)\over\chi(m_1,m_2)}\cr
B&~=~{{\vt_3}^4\over{\vt_2}^4}~{{\vt_3}^2\vt_3(m_1)\vt_3(m_2)-
{\vt_2}^2\vt_2(m_1)\vt_2(m_2)-{\vt_4}^2\vt_4(m_1)\vt_4(m_2)\over
\chi(m_1,m_2)} \ ,}}
integrate with respect to $u$, and then make the lowest order
approximation
$u=e^{-2\pi i\phi}$, we find the equation in \phidoneinst.

\goodbreak
\listrefs

\vfill
\eject
\end